\documentclass[epj]{svjour}
\usepackage{graphics}
\usepackage{graphicx}
\usepackage{enumerate}
\begin{document}
\title{Strong coupling effects in near-barrier heavy-ion elastic scattering}
\author{N. Keeley\inst{1} \and K.W. Kemper\inst{2,3} \and K. Rusek\inst{3}
}                     
\institute{National Centre for Nuclear Research, ul.\ Andrzeja So\l tana 7, 05-400 Otwock, Poland 
\and Department of Physics, The Florida State University, Tallahassee, Florida 32306, U.S.A. 
\and Heavy Ion Laboratory, University of Warsaw, ul.\ Pasteura 5a, 02-093 Warsaw, Poland}
\date{Received: date / Revised version: date}
%
\abstract{Accurate elastic scattering angular distribution data measured at bombarding energies just above 
the Coulomb barrier have shapes that can markedly differ from or be the same as the expected classical Fresnel scattering 
pattern depending on the structure of the projectile, the target or both. Examples are given such as $^{18}$O+$^{184}$W 
and $^{16}$O+$^{148,152}$Sm where the expected rise above Rutherford scattering due to 
Coulomb-nuclear interference is damped by coupling to the 
target excited states, and the extreme case of $^{11}$Li scattering, where coupling to the $^9$Li + $n$ + $n$ continuum leads to an 
elastic scattering shape that cannot be reproduced by any standard optical model parameter set. An early indication 
that the projectile structure can modify the elastic scattering angular distribution was the large vector analyzing 
powers observed in polarised $^6$Li scattering. The recent availability of high quality $^6$He, $^{11}$Li and $^{11}$Be 
data provides further examples of the influence that coupling effects can have on elastic scattering. Conditions for 
strong projectile-target coupling effects are presented with special emphasis on the importance of the beam-target 
charge combination being large enough to bring about the strong coupling effects. Several measurements are proposed that 
can lead to further understanding of strong coupling effects by both inelastic excitation and nucleon transfer on
near-barrier elastic scattering.
A final note on the anomalous nature of $^8$B elastic scattering is presented as it possesses 
a more or less normal Fresnel scattering shape whereas one would {\em a priori} not expect this due to the very low
breakup threshold of $^8$B. The special nature of $^{11}$Li is presented as it is predicted that no matter how far 
above the Coulomb barrier the elastic scattering is measured, its shape will not appear as Fresnel like whereas the elastic
scattering of all other loosely bound nuclei studied to date should eventually do so as the incident energy is increased,  
making both $^8$B and $^{11}$Li truly ``exotic''.
} 
\authorrunning{N. Keeley, K.W. Kemper and K. Rusek}
\titlerunning{Strong coupling effects \ldots}
\maketitle
\section{Introduction}
\label{intro}
\begin{figure}
\includegraphics[width=\columnwidth,clip=]{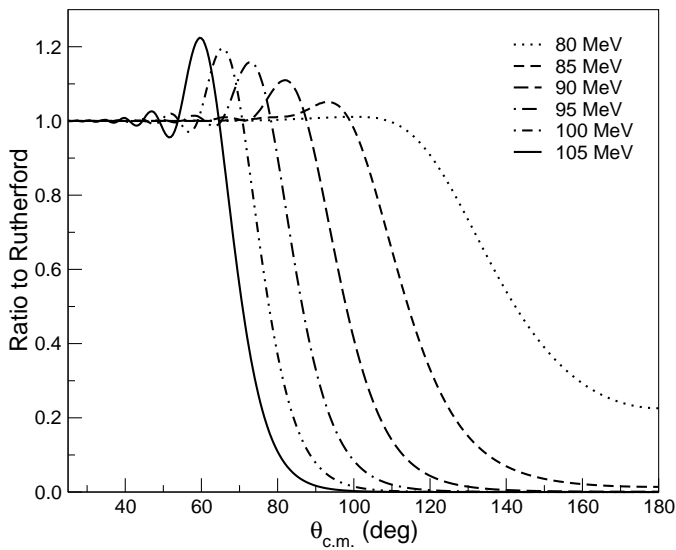}
\caption{\label{fig:ang-dist-e}Elastic scattering angular distributions for the $^{16}$O + $^{208}$Pb system  at several near-barrier
incident energies calculated using the optical model. Optical potential parameters were taken from a fit 
to the data of Ref.\ \cite{Rud01}. The nominal
Coulomb barrier for this system corresponds to an incident energy of about 83 MeV.} 
\end{figure}

Angular distributions of the differential cross section for heavy-ion elastic scattering at incident energies close to the Coulomb barrier,
when plotted as a ratio to the Rutherford scattering cross section, typically exhibit the characteristic
form often referred to as a Fresnel scattering pattern, viz.\ one or more minor oscillations about the Rutherford
value at small angles followed by a larger peak before an essentially exponential fall-off as a function of
scattering angle. The large peak is due to interference between scattering from the Coulomb and nuclear potentials and is
often referred to as the Coulomb-nuclear interference peak or the Coulomb rainbow peak. As the incident energy is reduced towards the 
Coulomb barrier the small-angle oscillatory structure and the Coulomb-nuclear interference peak move to larger 
angles and become less pronounced until they
disappear completely; Fig.\ \ref{fig:ang-dist-e} illustrates typical behaviour (optical model calculations are plotted rather than
actual data for the sake of clarity).

\begin{figure}
\includegraphics[width=\columnwidth,clip=]{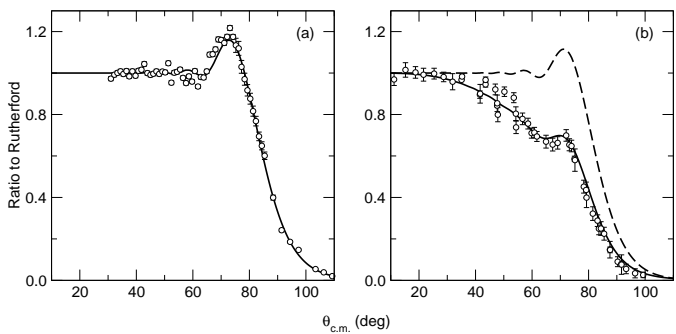}
\caption{\label{fig:ang-dist-comp}(a) Elastic scattering angular distribution for 95 MeV $^{16}$O incident on a $^{208}$Pb 
target. The circles represent the data of Ref.\ \cite{Rud01} (re-normalised to the Rutherford value at forward angles
on this plot) while the solid curve represents an optical model fit. (b) Elastic scattering angular distribution for  
90 MeV $^{18}$O incident on a $^{184}$W target. The circles represent the data of Ref.\ \cite{Tho77} while the solid
curve denotes the result of a coupled channels calculation including Coulomb and nuclear coupling to the  first $2^+$
excited state of $^{184}$W. The dashed curve denotes the result of a no coupling optical model calculation using the
same optical potential as in the coupled channels calculation.} 
\end{figure}

However, some systems are found to have angular distributions that differ markedly from the norm in that where the Coulomb-nuclear
interference peak should be there is instead a large {\em reduction} of the elastic scattering cross section compared
to the Rutherford value. Figure \ref{fig:ang-dist-comp} graphically illustrates this phenomenon: Fig.\ \ref{fig:ang-dist-comp} (a)
shows a normal near-barrier heavy-ion elastic scattering angular distribution, that for 95 MeV $^{16}$O incident on a $^{208}$Pb
target \cite{Rud01}, which contrasts with that in Fig.\ \ref{fig:ang-dist-comp} (b) for 90 MeV $^{18}$O incident on a $^{184}$W
target \cite{Tho77}. 
The observed depletion of the elastic scattering cross section in the $^{18}$O + $^{184}$W system 
is due to strong coupling, in this case strong quadrupole Coulomb coupling to the first $2^+$ excited state 
of $^{184}$W \cite{Tho77}. Similar effects have also been observed due to strong quadrupole Coulomb coupling
in the projectile in the $^{20}$Ne + $^{208}$Pb system \cite{Gro78}.
It is these strong coupling effects on the elastic scattering angular distribution that form the subject of this review.
Such a review is timely since with the increasing quality of radioactive beams rather precise measurements of angular
distributions of the elastic scattering of exotic nuclei are now feasible and it is possible to probe the projectile
structure through such scattering. Such measurements have already been made for several
weakly-bound exotic nuclei and the resulting angular distributions are found to exhibit all the characteristics of
strong coupling effects, in these cases due to strong coupling to the projectile continuum. 

Note that in both figures we have plotted the cross section on a linear scale rather than the more usual logarithmic one.
In this we follow the recommendation of Satchler \cite{Sat75,Bal75} who pointed out that a semi-logarithmic plot can
obscure the details of the nuclear-Coulomb interference region, the part of the angular distribution most sensitive
to the nuclear interaction:
``A particularly sensitive part of the angular distribution is the oscillatory region just before the exponential fall
below the Rutherford cross section. (The details of this region may be overlooked if, as is often done, the ratio-to-Rutherford
cross section is shown on a semi-logarithmic plot. It is more revealing to use a linear plot.)'' \cite{Bal75}.

We begin with a survey of elastic scattering measurements for systems involving stable nuclei that exhibit strong coupling 
effects from both inelastic and transfer couplings. This survey aims to be representative rather than exhaustive, 
in that we give examples of the types of system
that exhibit strong coupling effects. We then review measurements of the elastic scattering of exotic nuclei that also
exhibit strong coupling effects before summarising the necessary conditions for the existence of these effects. Finally
we make some suggestions for future measurements along with the accuracy required to explore further this phenomenon.  

\section{Strong coupling effects with stable beams}
\subsection{Systems with effects due to target coupling}
\label{targets}
\begin{figure}
\includegraphics[width=\columnwidth,clip=]{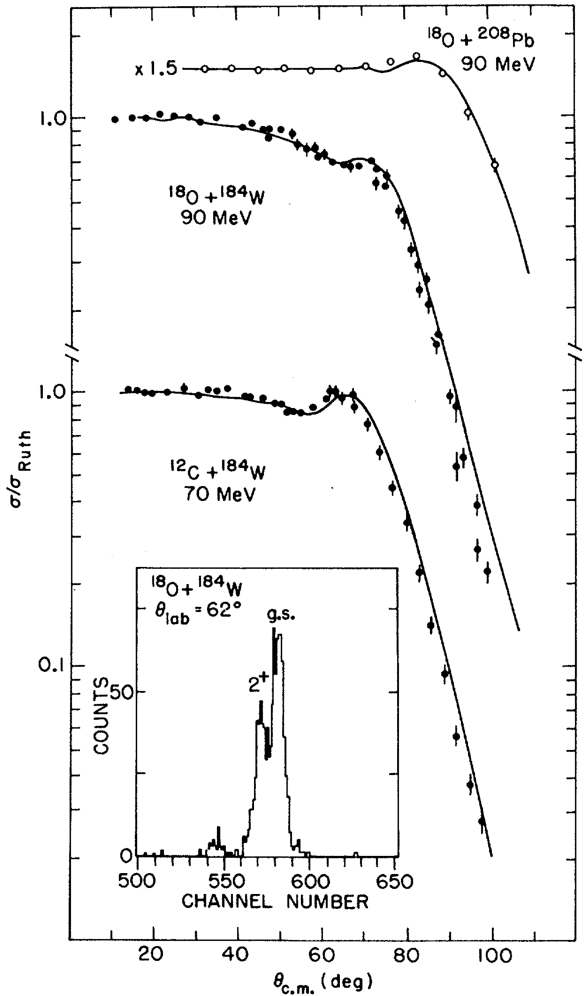}
\caption{\label{fig:OW}Elastic scattering data for the $^{18}$O + $^{184}$W and $^{12}$C + $^{184}$W systems at 90 and 70 MeV
respectively. The solid curves denote the results of optical model ($^{18}$O + $^{208}$Pb) or coupled channels ($^{18}$O
+ $^{184}$W and $^{12}$C + $^{184}$W) calculations. Taken from Ref.\ \cite{Tho77}. Reprinted Fig.\ 1 with 
permission from \cite{Tho77}. \textcircled{c} 1977, The American Physical Society}
\end{figure}

The first published data to show strong coupling effects on the elastic scattering were for the $^{18}$O and $^{12}$C 
+ $^{184}$W systems at incident energies of 90 and 70 MeV respectively \cite{Tho77}. As can be seen in Fig.\ \ref{fig:OW}
the effect is less marked for the $^{12}$C + $^{184}$W data, due to the smaller charge product for this projectile,
although the $^{12}$C incident energy is somewhat higher with respect to the Coulomb barrier than for the $^{18}$O + 
$^{184}$W data and this will also play a r\^ole in reducing the effect.
Also shown in Fig.\ \ref{fig:OW} is the measured angular distribution for the elastic scattering of a 90 MeV $^{18}$O beam
from a $^{208}$Pb target which exhibits the usual Fresnel shape, providing further proof that the strong coupling effect
is due to the $^{184}$W target and not the projectile.
Incidentally, a comparison of Fig.\ \ref{fig:ang-dist-comp} and  Fig.\ \ref{fig:OW} well illustrates the value of 
presenting near-barrier heavy-ion elastic scattering angular distributions on a linear scale.  

These data were taken with the use of a QDDD magnetic spectrometer in order to obtain the resolution necessary to
separate inelastic scattering to the 111 keV $2_1^+$ first excited state of $^{184}$W from the elastic scattering.
As Thorn et al.\ \cite{Tho77} were careful to point out, previous measurements of the ``elastic scattering'' of heavy
ions from heavy deformed targets obtained the usual Fresnel scattering angular distributions because they could not
resolve inelastic scattering to the ground state rotational bands. Thorn et al.\ also showed, by explicit coupled channels
calculations, that the elastic scattering data can be well described by coupling to the 111 keV $2_1^+$ state of $^{184}$W,
the effect on the elastic scattering being almost entirely due to the Coulomb coupling (it should be noted that these elastic
scattering data cannot be described by conventional optical model potentials).  

\begin{figure}
\includegraphics[width=\columnwidth,clip=]{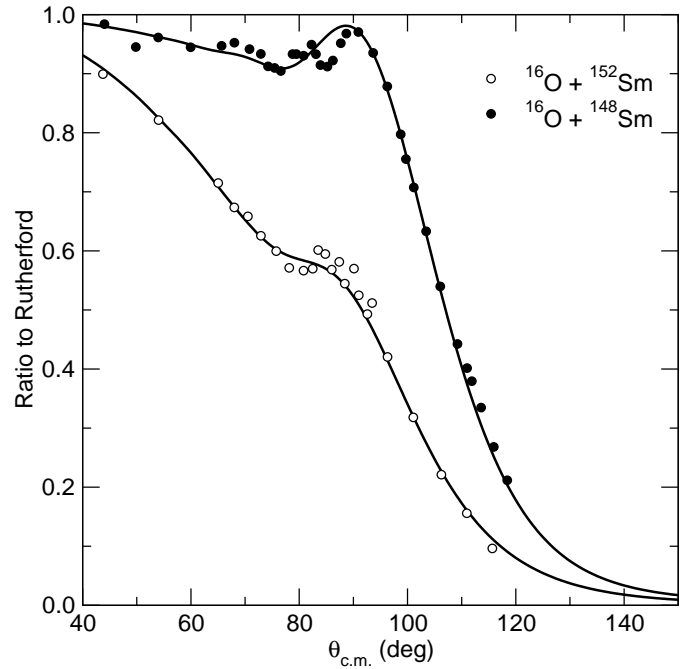}
\caption{\label{fig:O+Sm}Elastic scattering data for the $^{16}$O + $^{148,152}$Sm systems at incident energies of
72 MeV. The data are taken from Ref.\ \cite{Sto81}. The solid curves denote the result of coupled channels calculations
including $0^+ \longleftrightarrow 2^+ \longleftrightarrow 4^+$ rotational model coupling.} 
\end{figure}

Angular distributions for the elastic scattering of 72 MeV $^{16}$O ions from $^{152}$Sm \cite{Kim79,Sto81} and, to a  
lesser extent from $^{148}$Sm \cite{Sto81} show similar behaviour, see Fig.\ \ref{fig:O+Sm}.
Again, the main source of the effects on the elastic scattering angular distribution is Coulomb excitation of the first
$2^+$ excited states of the targets, although as Kim points out \cite{Kim79}, nuclear couplings are important for
a detailed reproduction of both the elastic and inelastic scattering angular distributions. Data were also obtained
at lower energies for the $^{16}$O + $^{148,150,152}$Sm systems by Talon {\em et al.\/} \cite{Tal81}. These data too
show significant deviations from Rutherford scattering but the effect on the shape of the angular distribution is not
so dramatic as for the higher energy data. Details of the experimental procedure for the 72 MeV $^{16}$O + $^{148,152}$Sm
measurements are not given in Refs.\ \cite{Kim79,Sto81} but the lower energy measurements of Talon {\em et al.\/} \cite{Tal81}
employed a QDDD magnetic spectrometer.

Figure \ref{fig:O+Sm} provides an unambiguous demonstration of the link between the importance of the coupling effect and the
strength of the $B(E2)$ and the excitation energy of the $2^+_1$ state since the charge products and, to a very
good approximation, the Coulomb barriers are identical for the $^{16}$O + $^{148}$Sm and $^{152}$Sm systems. While both these
Sm isotopes are deformed, $^{152}$Sm has a $B(E2; 0^+_1 \rightarrow 2^+_1)$ nearly five times that of $^{148}$Sm  plus a
much lower excitation energy. This is confirmed by the 72.3 MeV $^{16}$O + $^{144}$Sm elastic scattering data of Abriola
{\em et al.\/} \cite{Abr89} which show a classic Fresnel scattering angular distribution, $^{144}$Sm being a weakly-coupled
spherical nucleus with a relatively high-lying first excited state.  

The above examples demonstrate that, under the right conditions, strong coupling to low-lying excited states of the target
can give rise to near-barrier elastic scattering angular distributions that deviate significantly from the usual Fresnel
pattern. We shall explore fully what these conditions are in a later section. For the present, suffice it to say that
all systems where the effect has been observed and has been ascribed to target couplings involve heavy deformed target nuclei.  

\subsection{Systems with effects due to projectile coupling}
\label{projectiles}
\begin{figure}
\includegraphics[width=\columnwidth,clip=]{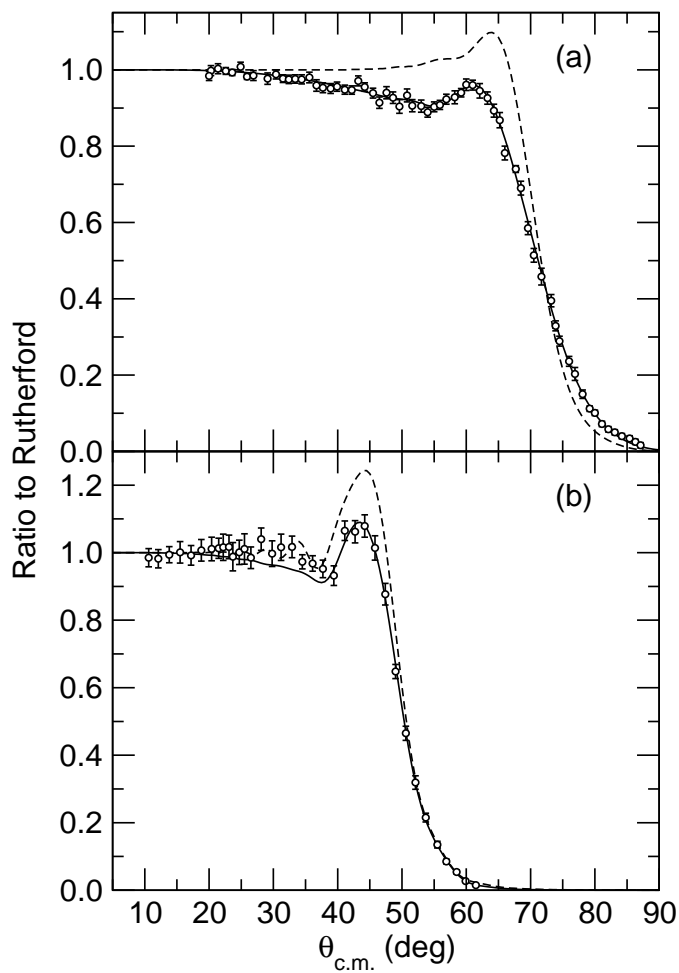}
\caption{\label{fig:Ne+Pb}Elastic scattering data for the $^{20}$Ne + $^{208}$Pb system at incident energies of
131 MeV (a) and 161.2 MeV (b). Data are taken from Refs.\ \cite{Gro78} and \cite{Bal75} respectively. The solid
curves denote the results of coupled channels calculations including coupling to the $2^+_1$ state of $^{20}$Ne.
The dashed curves denote the results of no coupling optical model calculations using the same
optical potentials as in the appropriate coupled channels calculations.}
\end{figure}

Elastic scattering angular distributions for the $^{20}$Ne + $^{208}$Pb system have been measured for incident $^{20}$Ne
energies of 161.2 MeV \cite{Bal75} and 131 MeV \cite{Gro78}. The 161.2 MeV data at first sight appear to show the usual
Fresnel scattering pattern and an acceptable description of these data could be obtained with standard optical potentials,
with the exception of the angular region between $34^\circ$ and $50^\circ$ \cite{Bal75}. The 131 MeV data,
however, unambiguously show the strong coupling effect, the shape of the angular distribution being similar to that for
the $^{16}$O + $^{148}$Sm system (cf.\ Figs.\ \ref{fig:O+Sm} and \ref{fig:Ne+Pb}).  
Both data sets were obtained using position sensitive silicon detectors mounted on moveable arms.
Coupled channels calculations reveal that the dip in the elastic scattering angular distribution just before the main
Coulomb-nuclear interference peak at 161.2 MeV is a residual effect due to the strong coupling to the $^{20}$Ne $2^+_1$
state, see Fig.\ \ref{fig:Ne+Pb} (b). These data therefore clearly show that the strong coupling effect diminishes as
the incident energy is increased. 

\begin{figure}
\includegraphics[width=\columnwidth,clip=]{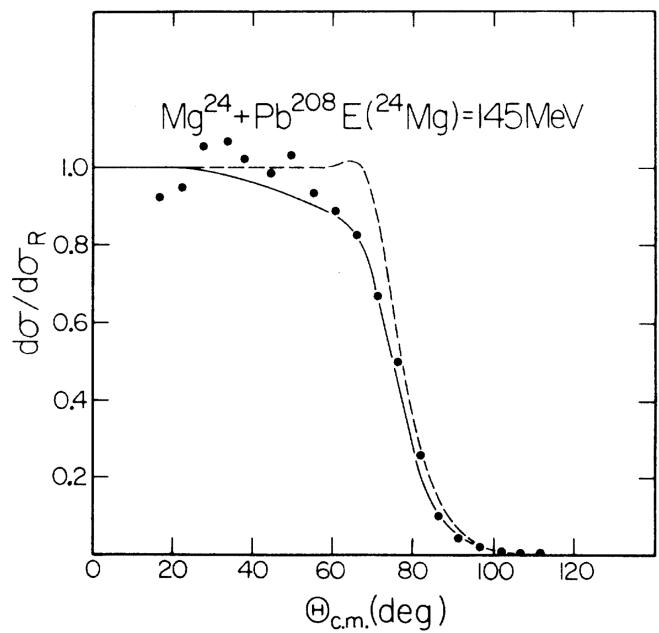}
\caption{\label{fig:Mg+Pb}Elastic scattering data for the $^{24}$Mg + $^{208}$Pb system at an incident energy of
145 MeV. The solid curve denotes an optical model fit including a long-range polarisation potential due to Coulomb
excitation of the $2^+_1$ of $^{24}$Mg while the dashed curve indicates the result of an optical model calculation
with the same potential but without the Coulomb polarisation potential.  Taken from Ref.\ \cite{Eck81}. Reprinted Fig.\ 4 with
permission from \cite{Eck81}. \textcircled{c} 1981, The American Physical Society}
\end{figure}

Elastic scattering angular distributions have also been measured for the $^{24}$Mg + $^{208}$Pb and $^{26}$Mg + $^{208}$Pb
systems at incident energies of 145 MeV \cite{Eck81} and 200 MeV \cite{Hen89} and 200 MeV \cite{Hen91} respectively.
The 145 MeV $^{24}$Mg + $^{208}$Pb data were taken using an Enge split pole magnetic spectrometer but both sets of 200
MeV data employed standard position sensitive silicon detectors. The 200 MeV data have elastic scattering angular
distributions of similar shape to that for the 161.2 MeV $^{20}$Ne + $^{208}$Pb data, i.e.\ there is only a residual strong
coupling effect, this time clearly visible as a distinct dip in the cross section just before the main Coulomb-nuclear
interference peak. The effect is smaller for $^{26}$Mg, as might be expected due to its weaker collectivity compared
to $^{24}$Mg. The 145 MeV $^{24}$Mg + $^{208}$Pb data clearly show the strong coupling effect, see Fig.\ \ref{fig:Mg+Pb}.

Our final example of a system with strong coupling effects due to the projectile is $^{28}$Si + $^{208}$Pb. Elastic
scattering data for this system are available at incident energies of 162 MeV \cite{Eck81}, 166 MeV \cite{Voj87},
209.8 MeV \cite{Chr84} and 225 MeV \cite{Kol84}. The influence of strong coupling is apparent for all these data
sets although the effect becomes weaker as the incident energy is increased. The measurements of Refs.\ \cite{Eck81,Voj87,Kol84}
were made using Enge split pole spectrometers while that of Ref.\ \cite{Chr84} employed a QDDD spectrometer. 

For all these systems it is strong Coulomb coupling to the $2^+_1$ excited state of the projectile that is predominantly  
responsible for the strong coupling effect on the elastic scattering, similarly to what was found for the systems with
strongly coupled targets. This raises the intriguing question as to what the angular distribution for a system with a
strongly deformed projectile {\em and} a strongly deformed target would look like.  

\subsection{Systems with effects due to projectile and target coupling}
\label{proj-targ}
\begin{figure}
\includegraphics[width=\columnwidth,clip=]{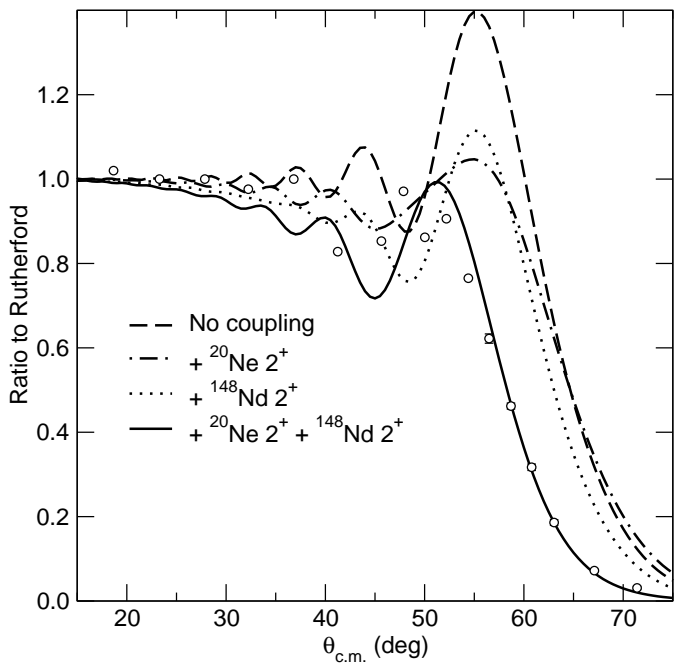}
\caption{\label{fig:Ne+Nd}Elastic scattering data for the $^{20}$Ne + $^{148}$Nd system at an incident energy of
116 MeV. The data are taken from Ref.\ \cite{Jia91}. The solid curve denotes the result of a coupled channels calculation
including both $^{20}$Ne and $^{148}$Nd $0^+_1 \rightarrow 2^+_1$ coupling (but no mutual excitation), the dotted curve
the result of a calculation including the $^{148}$Nd $0^+_1 \rightarrow 2^+_1$ coupling only, the dot-dashed curve a
calculation including the $^{20}$Ne  $0^+_1 \rightarrow 2^+_1$ coupling only and the dashed curve the no coupling
calculation.}
\end{figure}

As an example of a system with strong coupling in both projectile and target we cite $^{20}$Ne + $^{148}$Nd.
Data for the elastic and inelastic scattering in this system  were measured using a QDDD
magnetic spectrometer at an incident energy of 116 MeV \cite{Jia91}. 
The shape of the angular distribution is similar to those for systems with strong
coupling in either the target alone or the projectile alone, thus the presence of strong coupling in both target and
projectile does not seem to affect the qualitative nature of its effect on the elastic scattering, see Fig.\ \ref{fig:Ne+Nd}. 

In spite of a large cross section for mutual excitation of the $^{20}$Ne and $^{148}$Nd $2^+_1$ states it was found that
coupling to mutual excitation had little influence on the elastic scattering angular distribution \cite{Jia91}. The effect seems to
be due to the individual couplings to the separate $2^+_1$ states, although the calculations plotted in Fig.\ \ref{fig:Ne+Nd}
suggest that the combined effect of projectile and target couplings is greater than the sum of its parts.

\subsection{Strong coupling effects with polarised projectiles}

An example of a different type of strong coupling effect is provided by the analysing powers for elastic scattering
of polarised heavy ion beams. 
Analysing powers measured in experiments with polarised beams were found to be even more sensitive to the coupling effects than the
cross sections. In the 1980s many experiments with polarised $^{6,7}$Li and $^{23}$Na beams were performed at the 
Max Planck Institute for Nuclear Physics in Heidelberg, Germany~\cite{fick}. Later on these studies were moved to the Nuclear Structure
Facility of the Daresbury Laboratory, U.K., and in the second part of the 1990s to the Florida State University, U.S.A. The very first 
experiments with vector polarised $^6$Li beams scattered from nickel targets already
revealed large values of the vector analysing powers (iT$_{11}$) that could not be explained by a ``static'' spin-orbit interaction
derived from the deuteron-target or the nucleon-nucleon spin-orbit potentials by means of folding methods. The analysing powers 
predicted by optical model calculations with spin-orbit interactions calculated in this way were much too small to explain the 
experimental data.

\begin{figure}
\includegraphics[width=\columnwidth,clip=]{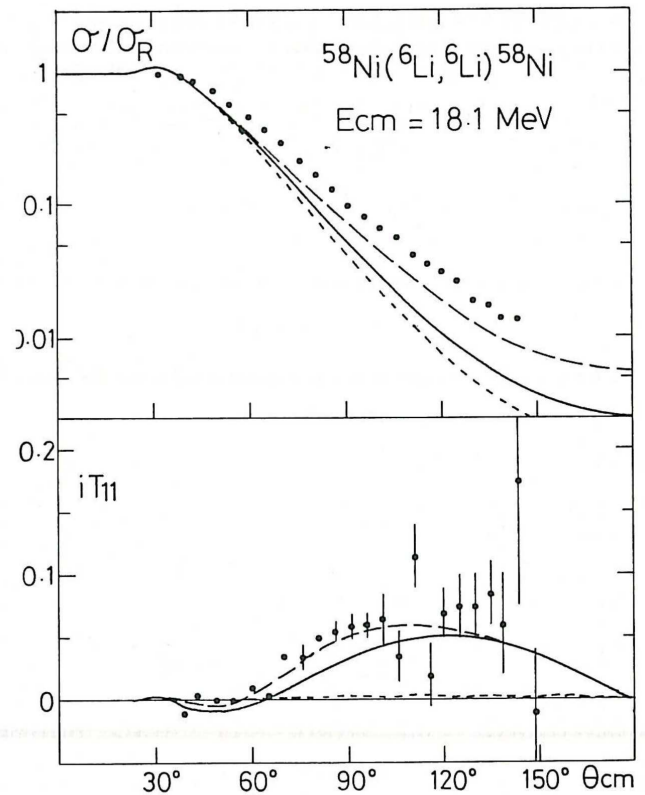}
\caption{\label{fig:nishioka}Elastic scattering angular distributions for polarised $^6$Li + $^{58}$Ni
compared to optical model and coupled channels calculations.
The analysing powers are almost completely generated by the coupling to the 3$^+$ resonant state of $^6$Li.  
Taken from Ref.\ \cite{nishioka}, Fig.\ 5.} 
\end{figure}

The explanation of this ``spin crisis'' came from microscopic coupled channels calculations that included couplings to the 
3$^+$ resonant first excited state of $^6$Li ~\cite{nishioka}. In Fig.\ \ref{fig:nishioka} the 
$^6$Li + $^{58}$Ni elastic scattering data (angular distributions of the differential cross section and the vector 
analysing powers) taken from ~\cite{rusek} are compared with the model predictions. The short-dashed curves 
represent the results of an optical model calculation 
with the spin-orbit potential derived from the deuteron-target interaction by means of the 
Watanabe-type cluster-folding method. This calculation underpredicted the
measured differential cross section and generated values of the analysing power that are very close to zero. Inclusion of couplings 
to the resonant first excited state of the projectile in coupled channels calculations produced much larger 
values of the analysing powers, close to experiment
(solid curves). The inclusion of further couplings to the resonances at 4.31 and 5.65 MeV excitation energy improved the
description of the data still further (dashed curves). Further improvement would require couplings to the non-resonant
$\alpha$ + $d$ continuum.   

\begin{figure}
\includegraphics[width=\columnwidth,clip=]{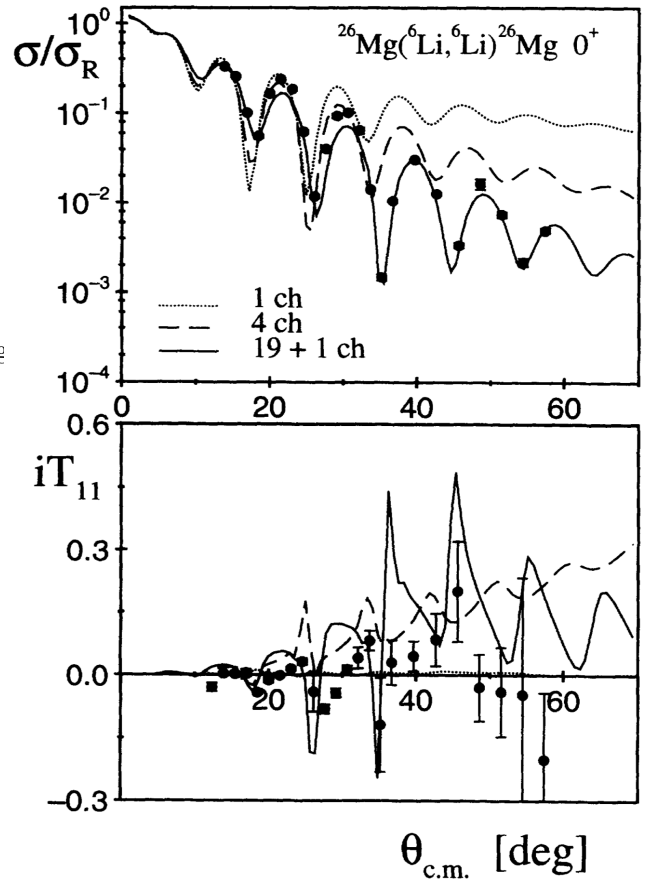}
\caption{\label{fig:rusek}Comparison of 60 MeV $^6$Li + $^{26}$Mg elastic scattering data with the results of coupled channels
calculations. The dotted curves denote the result of an optical model calculation including a cluster-folded spin-orbit
potential, the dashed curves denote the result of a coupled channels calculation including couplings to all three $^6$Li
resonant states and the solid curves denote the result of a coupled discretised continuum channels calculation including
couplings to the non-resonant $\alpha$ + $d$ continuum. Note that the analysing power arises entirely from channel couplings. 
Taken from Ref.\ \cite{rusek1}. Reprinted
Fig.\ 3 with permission from \cite{rusek1}. \textcircled{c} 1994, The American Physical Society} 
\end{figure}
This is illustrated in Fig.\ \ref{fig:rusek}. The experimental data for 60 MeV polarised $^6$Li + $^{26}$Mg elastic 
scattering were obtained at the Nuclear Structure Facility of the Daresbury 
Laboratory ~\cite{rusek1}. As in the previous case, optical model calculations with the cluster-folded spin-orbit potential 
included generated negligible values of the vector analysing power (dotted curves). Couplings to all three resonances of
$^6$Li improved the description of the differential cross section angular 
distribution and produced sizable values of the analysing power (dashed curves). Finally, inclusion of couplings to the non-resonant 
continuum resulted in a good reproduction of both sets of data (solid curves). One should mention that these calculations were free of any
adjustable parameters.

Finally, in Ref.\ \cite{Van89} an extensive coupled channels analysis of $^6$Li + $^{12}$C and $^6$Li + $^{16}$O elastic 
and inelastic scattering data found that for these light target systems the elastic scattering vector analysing power could
only be satisfactorily explained by explicitly including couplings to the $L = 2$ resonances of $^6$Li. However, 
it was found that the static spin-orbit potential did have an influence on the final result for the calculated
analysing power for these systems via a complicated interference with the dynamic coupling effects.  

\subsection{Strong coupling effects without strong couplings}
\label{paradox}
The title of this final sub-section dealing with strong coupling effects on elastic scattering involving stable heavy-ion
beams would appear at first sight to be something of a paradox: how is it possible to have a ``strong coupling'' effect on the elastic
scattering in the absence of strong couplings? However, this phenomenon does occur.
\begin{figure}
\includegraphics[width=\columnwidth,clip=]{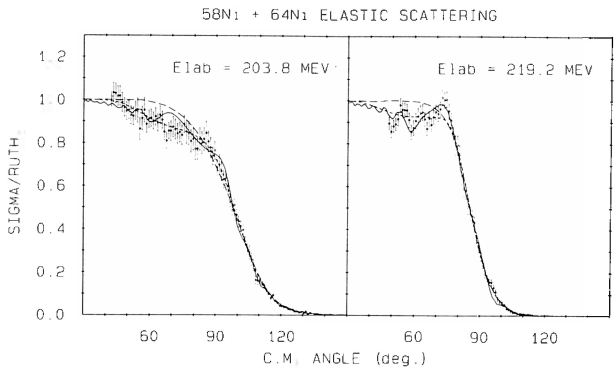}
\caption{\label{fig:Ni+Ni}Elastic scattering data for the $^{58}$Ni + $^{64}$Ni system at incident energies of
203.8 MeV (left) and 219.2 MeV (right). The solid curves denote coupled channels calculations including coupling to
the $2^+_1$ states of both nuclei while the dashed curves are optical model calculations. Taken from Ref.\ \cite{Rui92}, Fig.\ 4.}
\end{figure}
Neither $^{58}$Ni nor $^{64}$Ni are 
considered to exhibit strong coupling, their relatively high-lying $2^+_1$ levels ($E_{\mathrm{x}} = 1.45$ MeV for $^{58}$Ni and
1.35 MeV for $^{64}$Ni) being rather weakly-coupled vibrational states. The elastic scattering data of Ref.\ \cite{Rui92} for
the $^{58}$Ni + $^{64}$Ni system at two near-barrier energies nevertheless show the characteristic ``strong coupling'' form 
rather than the usual Fresnel scattering pattern, see Fig.\ \ref{fig:Ni+Ni}. The apparent paradox may be resolved by considering
the large charge product for this system which transforms the weak coupling of the Ni $2^+_1$ levels into a strong Coulomb
coupling. 

\begin{figure}
\includegraphics[width=\columnwidth,clip=]{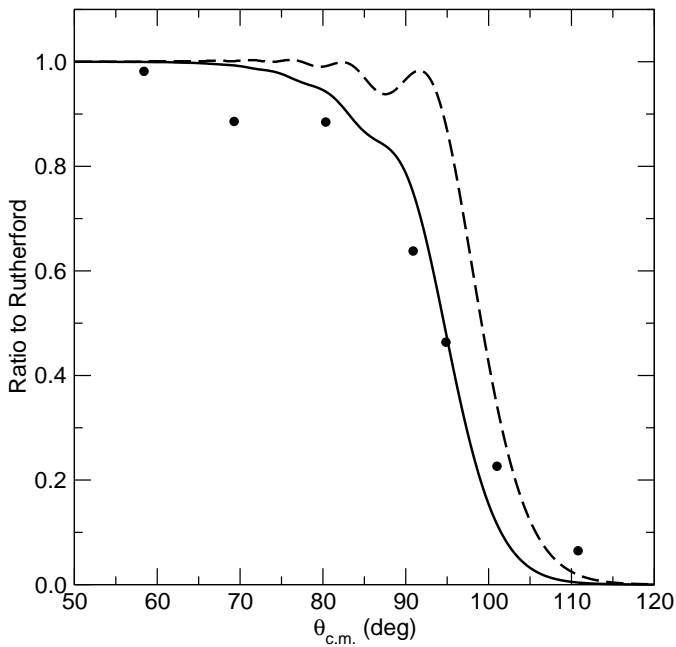}
\caption{\label{fig:40ca208pb}Calculated elastic scattering angular distributions for the $^{40}$Ca + $^{208}$Pb system 
at an incident energy of 236 MeV for the no coupling (dashed curve) and coupled reaction channels (solid curve) cases.
The coupled reaction channels calculation includes couplings to the $^{208}$Pb($^{40}$Ca,$^{41}$Ca)$^{207}$Pb single
neutron pickup reaction only. The points denote the quasi-elastic scattering data of Ref.\ \cite{Szi04}.}
\end{figure}

Similar effects should be exhibited by the near-barrier elastic scattering of most weakly-coupled medium mass vibrational nuclei
from a sufficiently heavy target. This indeed seems to be the case, see e.g.\ the $^{40}$Ar + $^{208}$Pb elastic scattering data
of Ref.\ \cite{Arl80} and the $^{58}$Ni + $^{208}$Pb data of Ref.\ \cite{Bec87}. However, a possible exception is the $^{40}$Ca
+ $^{208}$Pb system. A quasi-elastic scattering angular distribution for this system has been reported for an incident $^{40}$Ca
energy of 236 MeV \cite{Szi04}. While it is not explicitly stated in Ref.\ \cite{Szi04} exactly which channels are included in
the definition of ``quasi-elastic'' the angular distribution does exhibit the shape associated with strong coupling effects, 
from which we may infer that the elastic scattering angular distribution will also do so. However, coupled channels calculations
suggest that, due to the doubly-magic nature of $^{40}$Ca and its consequent structure properties, the effect is not produced
by inelastic couplings. Rather, it appears that this system may be a case where strong transfer couplings produce a similar
effect on the elastic scattering. Calculations presented in Ref.\ \cite{Szi04} seem to bear this out and conventional
coupled reaction channels calculations including the $^{208}$Pb($^{40}$Ca,$^{41}$Ca)$^{207}$Pb single neutron pickup coupling confirm this,
see Fig.\ \ref{fig:40ca208pb}.

\section{Strong coupling effects with radioactive beams}
\label{sec:RIBs}

The elastic scattering of certain light, weakly-bound radioactive nuclei from heavy targets has also been found to exhibit the
non-Fresnel pattern angular distributions characteristic of strong coupling effects. For these nuclei it is strong coupling to
the low-lying continuum that is responsible. A general review of available data for light radioactive beams was given in Ref.\
\cite{Kee09} so that we shall only give representative examples here.  

\begin{figure}
\includegraphics[width=\columnwidth,clip=]{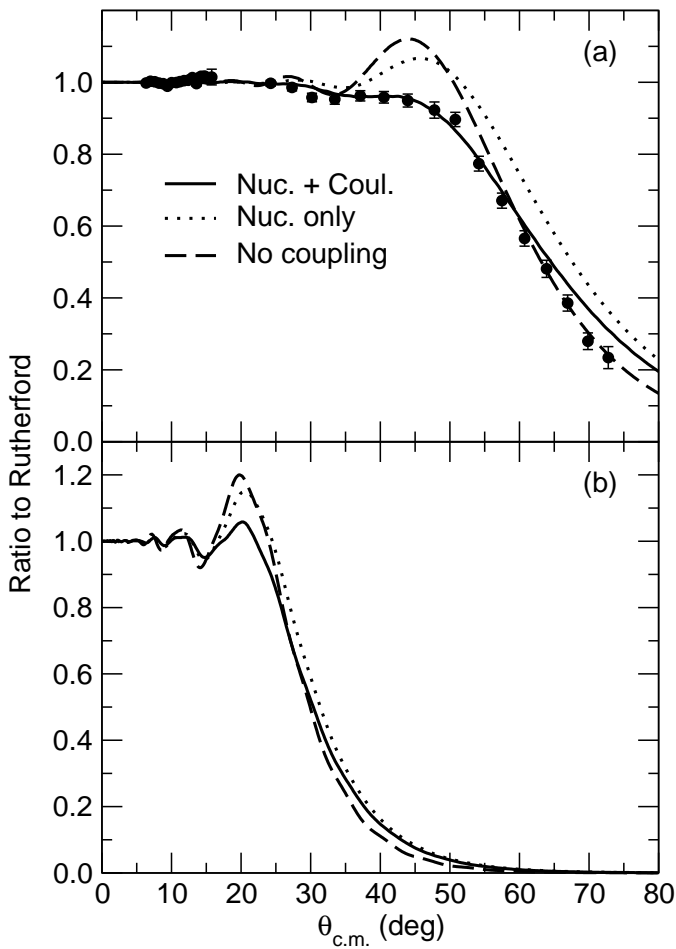}
\caption{\label{fig:6He}(a) Elastic scattering angular distribution for the $^{6}$He + $^{208}$Pb system at an incident energy of
27 MeV. The solid curve denotes a coupled discretised continuum channels calculation while the dashed curve denotes the corresponding
no coupling optical model calculation. The dotted curve denotes the result of a coupled discretised continuum channels calculation
with nuclear coupling only (but including diagonal Coulomb potentials). The data are taken from Ref. \cite{Kak03}. 
(b) The solid curve denotes a coupled
discretised continuum channels calculation of the $^{6}$He + $^{208}$Pb elastic scattering at an incident $^6$He energy of 45
MeV while the dashed curve denotes the corresponding no coupling optical model calculation. The dotted curve denotes the result of a coupled
discretised continuum channels calculation with nuclear coupling only (but including diagonal Coulomb potentials).} 
\end{figure}

A particularly good example of the effect of strong coupling to the low-lying continuum on the near-barrier elastic scattering
of a weakly-bound radioactive nucleus is the $^6$He + $^{208}$Pb system. Figure \ref{fig:6He} (a) shows the 27 MeV $^6$He +
$^{208}$Pb elastic scattering angular distribution of Kakuee {\em et al.\/} \cite{Kak03}. 
Coupled discretised continuum channels calculations have confirmed that the effect is dominated by Coulomb dipole
coupling to the low-lying $\alpha + n + n$ continuum.  
However, since this is a strong Coulomb coupling effect the charge product of the projectile-target system must be sufficiently
large for the effect to be observed; the $^6$He + $^{64}$Zn data of Refs.\ \cite{Dip03,Dip04} exhibit the classic Fresnel shape.
This question was examined further in Ref.\ \cite{Kuc09} where a series of coupled discretised continuum channels calculations
suggested that a target with an atomic number greater than about 80 was required in order unambiguously to see the strong
coupling effect. Figure \ref{fig:6He} (b) underlines that careful selection of the incident energy is also required if the
strong coupling effect is to be observed; if the energy is too high with respect to the nominal Coulomb barrier then even
for a $^{208}$Pb target the $^6$He elastic scattering is expected to present a more or less conventional Fresnel type shape (45 
MeV represents an incident energy approximately 2.5 times that equivalent to the nominal Coulomb barrier for the $^6$He +
$^{208}$Pb system). 

Preliminary data for the elastic scattering of $^8$He from a $^{208}$Pb target at an incident energy of 22 MeV have
recently been published \cite{Mar13}. They show a similar shape to the $^6$He + $^{208}$Pb data at the same incident
energy \cite{Aco11}, although given the higher breakup thresholds of $^8$He, combined with a considerably larger
spectroscopic factor for single-neutron removal, coupling to neutron stripping reactions may turn out to have
a more prominent r\^ole in $^8$He elastic scattering than in $^6$He where its effect is most important at large
angles, see e.g.\ Ref.\ \cite{Kee08}. 

\begin{figure}
\includegraphics[width=\columnwidth,clip=]{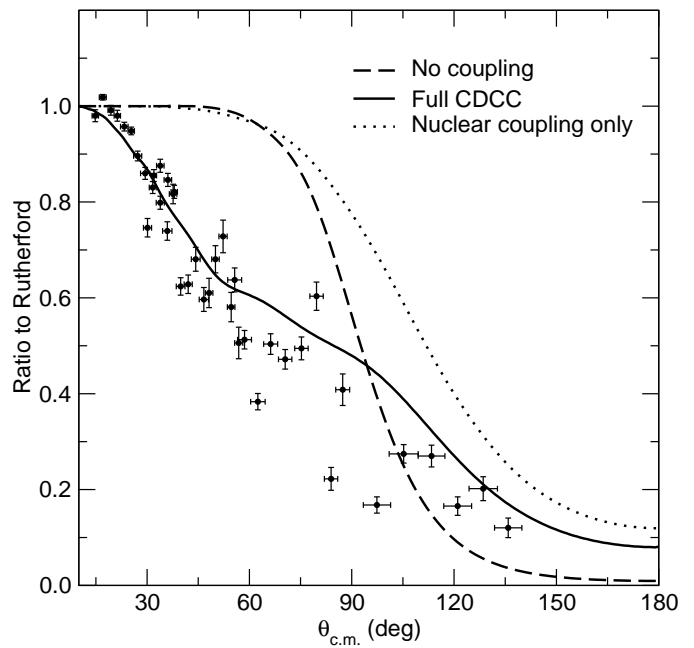}
\caption{\label{fig:11li}Elastic scattering angular distribution for the $^{11}$Li + $^{208}$Pb system at an incident energy of
29.8 MeV. The solid and dashed curves denote the results of coupled discretised continuum channels and no coupling 
calculations respectively. The dotted curve shows the result of a coupled discretised continuum channels calculation with
nuclear coupling only (but including diagonal Coulomb potentials). The data are taken from Ref.\ \cite{Cub12}.
Adapted from Ref.\ \cite{Kee13}, Fig.\ 1.}
\end{figure}

The elastic scattering of $^{11}$Li from $^{208}$Pb at near- and sub-barrier energies has recently been measured and 
shows spectacular departures from the Fresnel shape \cite{Cub12}. This too has been shown to be due mostly to strong Coulomb dipole
coupling to the low-lying continuum. The coupling effect is much stronger than for $^6$He, in part due to the extra charge
on the $^{11}$Li nucleus but mostly due to the much stronger Coulomb dipole polarisability of $^{11}$Li \cite{Kee13}.
Figure \ref{fig:11li} plots the data for 29.8 MeV $^{11}$Li + $^{208}$Pb elastic scattering compared to a coupled
discretised continuum channels calculation. The great importance of Coulomb couplings may be seen by comparing
the solid and dotted curves in the figure, nuclear coupling only becoming important for angles greater than about $80^\circ$.

\begin{figure}
\includegraphics[width=\columnwidth,clip=]{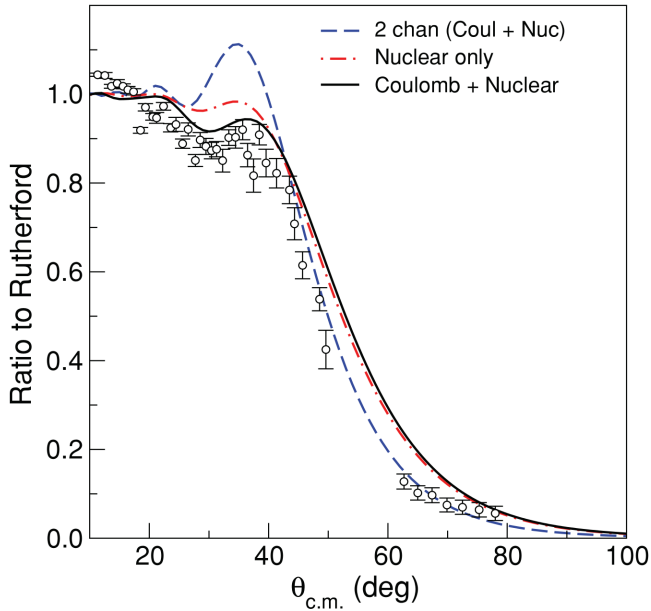}
\caption{\label{fig:11Be}Quasi-elastic scattering angular distribution for the $^{11}$Be + $^{64}$Zn system at an incident energy of
28.7 MeV. The solid curve denotes the result of the full coupled discretised continuum channels calculation including both
Coulomb and nuclear couplings, while the dot-dashed curve denotes the result of a similar calculation but with nuclear
couplings only while retaining the diagonal Coulomb potentials. The dashed curve is the result of a calculation including coupling
(both Coulomb and nuclear) to the 320 keV $1/2^-$ first excited state of $^{11}$Be only. Taken from Ref.\ \cite{Kee10}. Reprinted 
Fig.\ 3 with permission from \cite{Kee10}. \textcircled{c} 2010, The American Physical Society} 
\end{figure}

Quasi-elastic scattering data for the $^{11}$Be + $^{64}$Zn system have been measured at an incident energy of 28.7 MeV
\cite{DiP10}. The data are quasi-elastic since inelastic scattering to the low-lying bound first excited state of $^{11}$Be could
not be separated from the elastic scattering, although for most practical purposes they may be regarded as pure elastic. For this
system, and unlike the case for the $^6$He + $^{64}$Zn elastic scattering \cite{Dip03,Dip04}, the angular distribution has the
characteristic shape associated with strong coupling effects. The persistence of strong coupling effects for elastic
scattering from a medium mass target could partly be due to the extra charge on the $^{11}$Be nucleus compared to $^6$He.
However, coupled discretised continuum channels calculations found that, unlike all the other systems mentioned in this review,
strong {\em nuclear} couplings---in this case to the low-lying $^{10}$Be + $n$ continuum---appeared to have the most important
influence on the $^{11}$Be elastic scattering \cite{Kee10}, see Fig.\ \ref{fig:11Be}. 
The exact importance of the nuclear couplings is to some extent model dependent, see Ref.\ \cite{DiP12}. However, it
is clear that the nuclear couplings are far more important than for $^6$He or $^{11}$Li (or, indeed for the strongly-coupled
systems involving stable nuclei) and are vital in reproducing the shape of the angular distribution. The calculations of
Refs.\ \cite{Kee10} and \cite{DiP12} both neglect the possibility of excitation of the $^{10}$Be core. Extensions of the
standard coupled discretised continuum channels technique have been developed to include core excitation effects 
\cite{Sum06,Sum14,Die14} and have recently been applied to these data \cite{Die14}. It was found that the standard coupled
discretised continuum channels calculations of Ref.\ \cite{DiP12} gave similar results for the quasi-elastic scattering
to those of the more sophisticated model including core excitation \cite{Die14} although the latter was in better
agreement with the inclusive breakup cross section data.  

\begin{figure}
\includegraphics[width=\columnwidth,clip=]{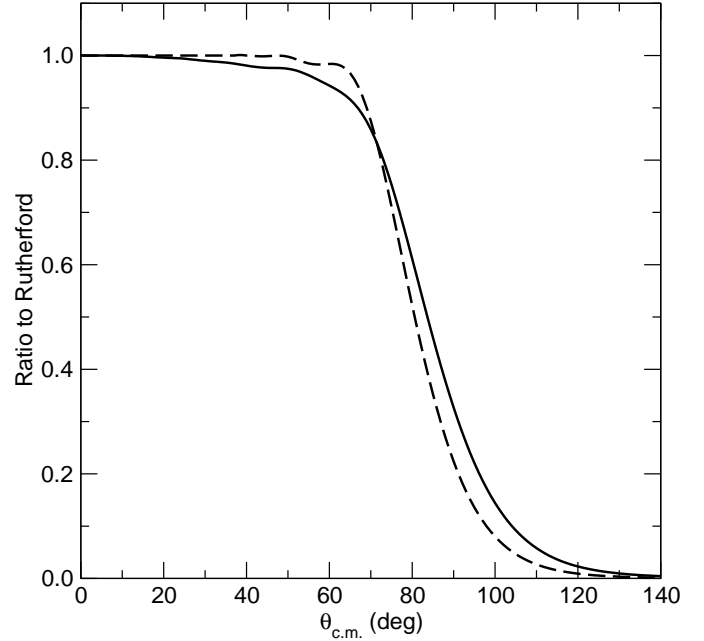}
\caption{\label{fig:8B}Elastic scattering angular distribution for the $^{8}$B + $^{208}$Pb system at an incident energy of
60 MeV. The full curve denotes the result of a coupled discretised continuum channels calculation while the dashed curve
represents the result of the no coupling optical model calculation. Adapted from Fig.\ 25 of Ref.\ \cite{Kee09} (see Ref.\
\cite{Kee09} for details of the calculation).}
\end{figure}

We end this brief survey of strong coupling effects for systems involving radioactive beams with another paradox. In this case
it is a nucleus that would be expected {\em a priori} to show such effects in its elastic scattering but which apparently does
not, $^8$B. The threshold against $^8$B $\rightarrow$ $^7$Be + $p$ breakup is just 137.5 keV so that the cross section for this
process should be large and one would naively expect commensurate effects on the elastic scattering due to coupling to the
continuum. However, coupled discretised continuum channels calculations suggest that the influence on the elastic scattering
of coupling to the $^7$Be + $p$ continuum is small, even for a $^{208}$Pb target \cite{Kee09}, see Fig.\ \ref{fig:8B}. 
While the model of $^8$B employed
in these calculations is simplified to the extent that excitations of the $^7$Be core are ignored, the 170.3 MeV 
$^8$B + $^{\mathrm{Nat}}$Pb elastic scattering data of Yang {\em et al.\/} \cite{Yan13} suggest that it is sufficiently
realistic. Coupling effects are expected to be minimal at this energy---approximately three times the Coulomb 
barrier---and indeed the measured angular distribution shows a classic Fresnel scattering pattern \cite{Yan13}. However,
calculations for near-barrier energies suggest that any suppression of the nuclear-Coulomb interference peak for
$^8$B + $^{208}$Pb elastic scattering will be due largely to ``static'' effects rather than coupling to the continuum \cite{Kee09}.
This is borne out by similar calculations \cite{Lub09} for the near-barrier $^8$B + $^{58}$Ni data of Ref.\ \cite{Agu09}.
Unfortunately $^8$B is a difficult beam to produce, particularly at lower energies, so that the available near-barrier
data do not have a sufficient number of points to confirm this experimentally \cite{Agu09}.

Part of the explanation for this behaviour may come from the fact that in $^8$B the ``valence'' particle is a proton rather
than a neutron, thus allowing subtle interference effects involving the Coulomb interaction with the core in addition
to the presence of a Coulomb barrier tending to suppress the large radius tail of the valence particle wave function. Indeed, 
Kumar and Bonaccorso \cite{Kum11} have shown that with regard to the nuclear breakup the proton in $^8$B 
behaves like a more tightly bound neutron. Later work showed the importance of interference effects for
breakup observables \cite{Kum12}. Nevertheless, the charge on the proton does not seem to account entirely for the lack
of coupling effect on the near-barrier $^8$B elastic scattering, since coupling to breakup has a much larger effect for $^6$Li 
elastic scattering at similar energies with respect to the Coulomb barrier in spite of the very much larger breakup
threshold (and consequently much smaller breakup cross section) for this projectile compared to $^8$B, see for example Ref.\
\cite{Kee10a}. 

\section{Conditions governing strong coupling effects}
\subsection{General considerations}

\begin{figure}
\includegraphics[width=\columnwidth,clip=]{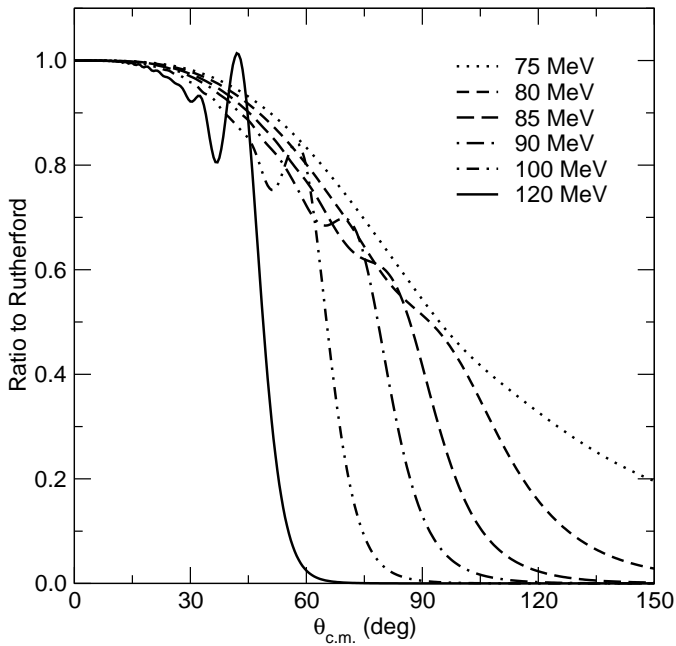}
\caption{\label{fig:s-c}Coupled channels calculations for the $^{18}$O + $^{184}$W system at several sub- and near-barrier
bombarding energies. The input parameters are identical with those of the coupled channels calculations in Ref.\ \cite{Tho77}.
The nominal Coulomb barrier for this system corresponds to a bombarding energy of about 77.7 MeV.}
\end{figure}

In this section we explore the conditions necessary for strong coupling effects on near-barrier elastic scattering
to manifest themselves. In the previous two sections we have seen how systems with strong coupling exhibit 
near-barrier elastic scattering angular distributions of a characteristic shape that, at first sight, appears to have nothing
in common with the usual Fresnel-type elastic scattering pattern. However, like the conventional Fresnel pattern the strong coupling
angular distribution shape does evolve as a function of bombarding energy, see for example the $^{28}$Si + $^{208}$Pb data of
Refs.\ \cite{Eck81,Voj87,Chr84,Kol84}. In order to illustrate this more clearly we present in Fig.\ \ref{fig:s-c} a series of
coupled channels calculations for the $^{18}$O + $^{184}$W system at incident energies ranging from somewhat below the nominal
Coulomb barrier ($V_{\mathrm{B}} = 70.8$ MeV, equivalent to a bombarding energy of about 77.7 MeV) to approximately $1.5 \times
V_{\mathrm{B}}$.  
The calculations use the same optical potential and coupling parameters as those of Ref.\ \cite{Tho77} since in this figure we
are investigating the qualitative evolution of the shape of the angular distribution as a function of bombarding energy rather
than attempting to make quantitatively accurate predictions of the elastic scattering at a particular energy.  

A comparison with Fig.\ \ref{fig:ang-dist-e} will show that, despite the superficially completely different shapes of the
elastic scattering angular distributions for the strongly-coupled system, what we find in Fig.\ \ref{fig:s-c} is in fact
a standard Fresnel-type scattering pattern superimposed on a strong Coulomb coupling angular distribution shape. 
As the bombarding energy is increased from 75 MeV we see an evolution from a completely structureless angular distribution
to the emergence of a strong Coulomb-nuclear interference peak as the real nuclear potential becomes of more importance,
exactly as in the classic Fresnel scattering pattern. The difference in the strongly-coupled system is that now, due to
depletion of the elastic scattering by the strong Coulomb excitation, the Coulomb-nuclear interference maximum is 
peaked at much less than the Rutherford value, except for energies considerably above the Coulomb barrier.

\subsection{Strong coupling equivalent to a long-range DPP}

Thus far we have explained strong coupling effects in terms of coupled channels calculations. However, on publication
of the $^{18}$O + $^{184}$W data of Thorn et al.\ \cite{Tho77} it was almost immediately demonstrated that they could
also be explained in terms of a long-range dynamic polarisation potential (DPP) induced by the Coulomb coupling which,
when added to a conventional optical model potential, provided a good description of the data. This DPP could
easily be calculated from a single parameter, the $B(E\lambda)$ value, making use of some approximations \cite{Lov77,Lov77a,Bal78,Bal79}.
The formalism of Baltz et al.\/ \cite{Bal78,Bal79} is more sophisticated than that of Love et al.\/ \cite{Lov77,Lov77a} although
both give similar results. There is now a considerable literature on the calculation using various approximations
of Coulomb excitation DPPs and their application to elastic scattering data, most recently including radioactive beam data.

Love et al.\ and Baltz et al.\ assume the sudden approximation, which leads to a predominantly imaginary 
DPP (the real parts of their DPPs are in fact negligible) and most subsequent work on Coulomb DPPs has made similar assumptions.
However, if the adiabatic approximation is appropriate, i.e.\ if the collision time $\gg$ the period of internal motion, a purely
real, attractive DPP results \cite{Sat83}. In reality, the DPP will be complex and the real and imaginary parts of the DPPs extracted 
from coupled discretised continuum channels calculations for light weakly-bound nuclei are usually comparable in size,
see e.g.\ Ref.\ \cite{Mac09}. The form of the real part of such DPPs at large radii matches very well the adiabatic
model expression for the (purely real) DPP which is directly related to the polarisability of the weakly-bound
nucleus. The dipole polarisability of light weakly-bound nuclei may thus be obtained from fits to the real part of the DPP   
extracted from appropriate coupled discretised continuum channels calculations, see for example Refs.\ \cite{Kee13} and \cite{Par11}.
Starting from analytic expressions for the imaginary polarisation potential for Coulomb dipole excitation, the corresponding real 
polarisation potential has been obtained through a dispersion relation involving the Coulomb dipole excitation energy \cite{And94},
thus yielding an analytic expression for a complex Coulomb dipole DPP. This formalism was applied to the $^{11}$Li + $^{208}$Pb
elastic scattering at 50 MeV, producing a prediction for the angular distribution in remarkable qualitative agreement with the  
recent measurements at somewhat lower incident $^{11}$Li energies \cite{Cub12}. 

While coupled channels calculations are now routine, at least for stable nuclei where the number of states involved is small,
the expressions for the Coulomb DPPs do give valuable insight into the conditions necessary for strong coupling
effects to be maximal. For example, Love et al.\/ \cite{Lov77a}, referring to their expression for the Coulomb DPP state that it:
``\ldots provides an explicit guide as to when coupling to a particular inelastic channel may be important. Clearly, the effects
should be most important in those cases in which the target and/or projectile has a large $B(E\lambda)$ and the projectile
and/or the target possesses a large Z.'' 

\subsection{The conditions under which strong coupling effects are expected}

This leads us conveniently into a discussion of the necessary conditions under which strong coupling effects on the near-barrier
elastic scattering are expected to appear.
Frahn and Hill \cite{Fra78} have given an excellent summary of how the strong
coupling influence depends on various parameters which we may conveniently quote here: 
``\ldots based solely on the long-ranged
nature of the polarization potential \ldots
some general predictions could be made:

\begin{enumerate}[i)]
\item{Dynamic polarization causes an overall reduction
of the differential cross section which is small at small
angles and becomes gradually stronger with increasing $\theta$.}

\item{The polarization probability increases rapidly
with increasing deformation of the target and/or
projectile.}

\item{For a given ratio of the energy to the Coulomb
barrier energy, the polarization increases strongly
with the mass numbers of projectile and target.}

\item{For a given projectile-target system the polarization
effect diminishes rapidly with increasing
energy.''}
\end{enumerate}

to which we might also add the excitation energy of the strongly coupled state; even a strongly coupled state will not produce
the characteristic strong coupling elastic scattering angular distribution if its excitation energy is high. This is, of
course, a somewhat subjective statement since whether or not the strong coupling effect is present also depends on the incident energy.
However, since such effects are usually limited to incident energies relatively close to the Coulomb barrier states with excitation
energies of approximately 5-10 MeV would not be expected to have much influence on the elastic scattering in the relevant energy regime
regardless of their coupling strength. A proper investigation of the interplay of these two linked properties---excitation energy of the
coupled state(s) and projectile incident energy---is beyond the scope of the present review but would form an interesting subject
for further study. 

Summarising the above information, we would expect strong coupling effects to be present in the near-barrier elastic scattering
under the following circumstances:

\begin{enumerate}
\item{Systems where the target is a well-deformed medium to heavy mass nucleus with a low-lying (a few hundred keV or less) 
first excited state. In this instance the projectile may be a relatively light, more or less inert nucleus such as $^{12}$C
or $^{16}$O or similar, the only requirement being that the charge product is sufficiently large since the Coulomb coupling
effect is dominant in such systems. If the target is sufficiently heavy, i.e.\ it carries a large enough charge, even $\alpha$
particle elastic scattering should show the strong coupling effect under these circumstances, see for example Refs.\ \cite{Lov77a,Rus09}.} 

\item{Systems where the projectile is a well-deformed light nucleus with a strongly coupled ground state rotational band and the
target is a heavy inert nucleus ($^{208}$Pb or similar). Again, since the coupling effect is Coulomb dominated exactly how heavy the
target nucleus needs to be for the effect to be plain will depend on the projectile charge. In this case the first excited state 
of the deformed projectile need not be particularly low-lying (1-2 MeV is sufficient).} 

\item{Systems where the projectile is a well-deformed light nucleus and the target a well-deformed medium to heavy mass nucleus.
It appears that coupling to mutual excitation of the target and projectile has little or no influence on the elastic scattering 
in these systems although the overall effect of individual coupling to the projectile and target excited states does seem to 
be greater than the sum of its parts.}

\item{Systems where both the projectile and target are of medium mass or greater. In this case the nuclei involved may have
only relatively weak couplings but the system will still exhibit strong coupling effects on the elastic scattering due to the
large charge product. Any medium mass (or greater) nucleus incident on, say, a $^{208}$Pb target should give rise to the
characteristic strong coupling shape for its near barrier elastic scattering. Doubly-magic medium mass nuclei may prove
exceptions, cf.\ $^{40}$Ca.}

\item{Systems with weakly-bound radioactive beams incident on heavy targets such as $^{197}$Au and $^{208}$Pb. There are two
caveats to this statement: a weak binding energy does not automatically guarantee the existence of the strong coupling
effect, as the case of $^8$B proves and a heavy target may not always be necessary to see the effect, cf.\ $^{11}$Be and
$^{11}$Li.} 

\item{Well angular momentum and Q value matched transfer reactions can give similar strong coupling effects to those due to
strong inelastic couplings. Such couplings may turn out to be important for exotic nuclei.} 
\end{enumerate}
In all these cases it should also be emphasised that the optimum effect will be for energies somewhat greater than the Coulomb
barrier, about 20--30 \% greater than $V_{\mathrm{B}}$. 

\begin{figure}
\includegraphics[width=\columnwidth,clip=]{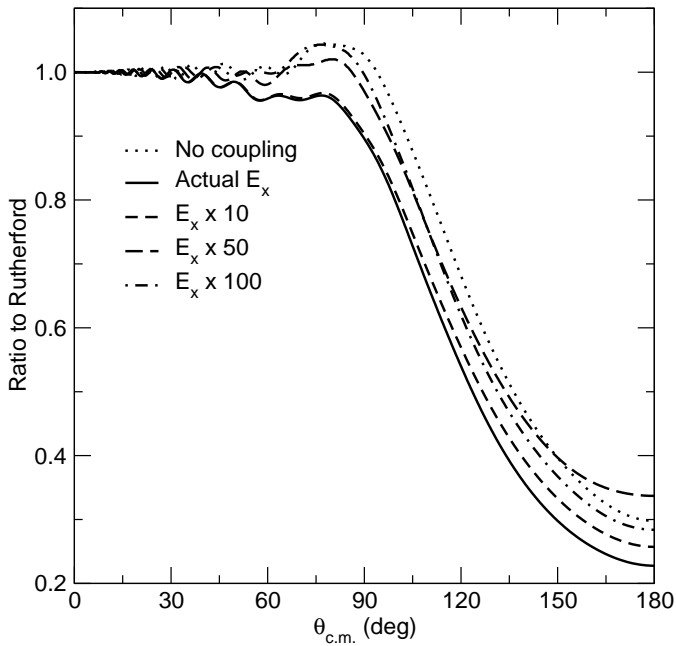}
\caption{\label{fig:a-U}Coupled channels calculations for the $\alpha$ + $^{238}$U system at a bombarding energy of 24.7 MeV.
The various curves correspond to a no coupling optical model calculation (dotted curve) and
coupled channels calculations with the excitation energy of the $^{238}$U $2^+_1$ state multiplied by factors of 1, 10, 50 and
100.} 
\end{figure}

Finally, to illustrate further the importance of excitation energy on the strong coupling effect at a given incident
energy close to the Coulomb barrier we present in Fig.\ \ref{fig:a-U}
test coupled channels calculations for the $\alpha$ + $^{238}$U  system at a bombarding energy of 24.7 MeV. The calculations
are similar to those of Rusek \cite{Rus09} except that for our purposes we have only included coupling to the $2^+_1$ state
of $^{238}$U (omission of coupling to the other members of the ground state band has little effect on the result beyond damping
out the small scale oscillations about the Rutherford value at small angles). 
The actual excitation energy of the $2^+_1$ state, 45 keV, was multiplied by factors of 10, 50 and 100 in order to investigate
the influence of excitation energy on the strong coupling effect. It is found that the excitation energy has to be multiplied by 
a factor of about 30, i.e.\ an excitation energy of 1.35 MeV, before the characteristic strong coupling shape has more or less 
disappeared in favour of the usual Fresnel-type elastic scattering pattern. Although there are still significant coupling effects
for $E_{\mathrm{x}} \times 50$ and 100 they are of such a nature that they could be subsumed into the optical model
fitting procedure. These calculations demonstrate that even a strongly coupled state will not give rise to the characteristic
strong coupling elastic scattering angular distribution shape if its excitation energy is too high, although exactly what 
constitutes ``too high'' may also depend on other factors such as the projectile incident energy.

\section{Suggested future experiments}

In this section we propose systems that would be interesting candidates for further study of the strong coupling effect in the
light of our conclusions concerning the conditions under which such effects should be most important. Suggestions are made for
experiments with both stable and radioactive beams and an indication of the required precision of the measurements is given.
We have limited ourselves to what we hope are practical measurements, either already or in the near future. All coupled
channels calculations in the following subsections were performed using the code FRESCO \cite{Tho88}. 

\subsection{Heavy beams on a heavy target}

Due to what we might term the magnifying effect of the large charge product, the elastic scattering of a heavy nucleus from a
similarly heavy target ought, in principle, to provide a good example of the ``strong coupling effect without strong coupling.'' 
As an example, we suggest the $^{120}$Sn + $^{208}$Pb system. Measurement of the pure elastic scattering would require a
detection energy resolution of about 1 MeV in order to resolve inelastic scattering to the first excited state of $^{120}$Sn 
($E_{\mathrm{x}} = 1.17$ MeV). An incident $^{120}$Sn energy of 900 MeV should be suitable, thus a detection energy resolution
of about 0.1 \% would be required.

\begin{figure}
\includegraphics[width=\columnwidth,clip=]{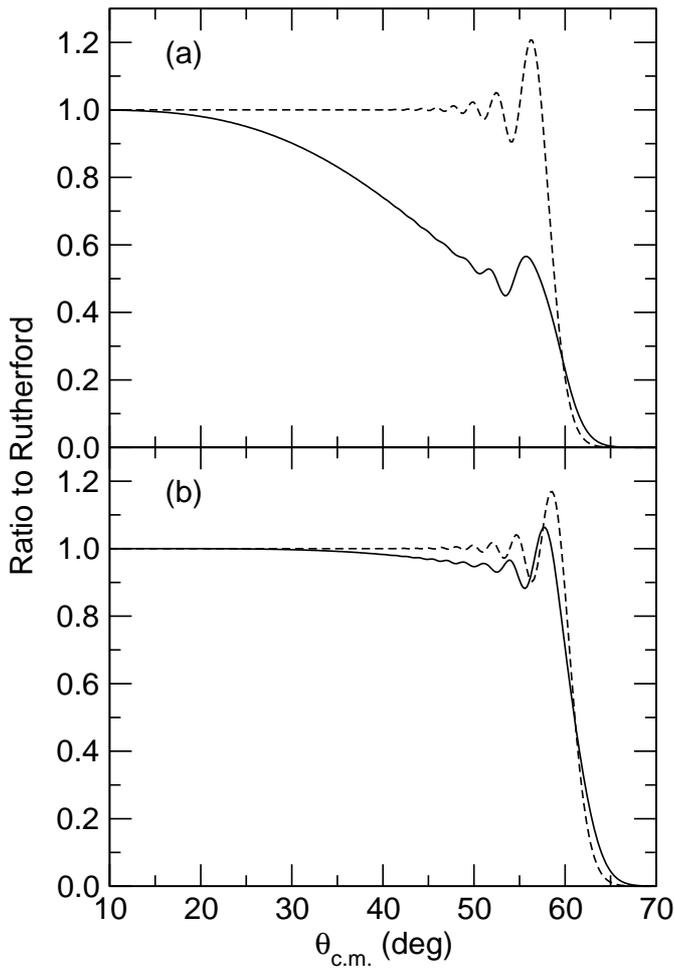}
\caption{\label{fig:SnPb}(a) Coupled channels calculation for the $^{120}$Sn + $^{208}$Pb system at a bombarding energy of 900 MeV.
The calculation including Coulomb and nuclear couplings to the $2^+_1$ state of $^{120}$Sn and the $3^-_1$ state of $^{208}$Pb 
is denoted by the solid curve while the dashed curve denotes the result of the no coupling optical model calculation.
(b) Coupled channels calculation for the $^{132}$Sn + $^{208}$Pb system at a bombarding energy of 900 MeV.
The calculation including Coulomb and nuclear couplings to the $2^+_1$ state of $^{132}$Sn and the $3^-_1$ state of $^{208}$Pb
is denoted by the solid curve while the dashed curve denotes the result of the no coupling optical model calculation.} 
\end{figure}

We present the result of a coupled channels calculation for a 900 MeV $^{120}$Sn beam incident on a $^{208}$Pb target in
Fig.\ \ref{fig:SnPb} (a). 
The calculation does not aim at a quantitatively accurate prediction of the angular distribution but it should be reasonably
realistic. The optical potential was calculated according to the prescription of Broglia and Winther \cite{Bro81} with
the imaginary part taken to be the same as the real part but with a depth divided by a factor of 4. The $B(E2)$ and $B(E3)$
values were taken from Raman et al.\/ \cite{Ram01} and Kib\'edi and Spear \cite{Kib02} respectively. The nuclear deformation
lengths were derived from the $B(E\lambda)$ values assuming the collective model and radii of $1.2 \times \mathrm{A}^{1/3}$ fm. 
A precision of $\pm 5$ \% on each point of the angular distribution would be more than sufficient not only to demonstrate
the effect but also show the details of the structure, provided that a sufficient number of points could be measured. Such
a measurement should therefore be feasible over a reasonable period of time given the availability of a suitable $^{120}$Sn
beam. 

By way of contrast, in Fig.\ \ref{fig:SnPb} (b) we present the result of a coupled channels calculation for the radioactive
$^{132}$Sn beam incident on a $^{208}$Pb target, again at an incident energy of 900 MeV. The $^{132}$Sn nucleus is doubly-magic
and its first excited state, a $2^+$ level, is consequently situated at the rather high energy of 4.04 MeV. We would therefore
anticipate a much reduced strong coupling effect. The calculation suggests that this will indeed be the case,
at least as far as inelastic excitations are concerned; in fact, the residual ``strong coupling'' effect is almost
entirely due to coupling to the $^{208}$Pb $3^-_1$ state. The coupled channels calculation took the $B(E2)$ value for
$^{132}$Sn from Pritychenko et al.\ \cite{Pri13}, the nuclear deformation length being derived in a similar fashion to those
in the $^{120}$Sn + $^{208}$Pb calculation. The optical potential was again calculated using the prescription of Broglia and
Winther \cite{Bro81}. In this case a detection energy resolution of about 0.25 \% would be required, since it would be necessary
to resolve the $3^-_1$ state of $^{208}$Pb ($E_{\mathrm{x}} = 2.61$ MeV) in order to measure pure elastic scattering. A
precision of $\pm 5$ \% on each point of the angular distribution would be more than adequate to demonstrate the
predicted difference between the $^{120}$Sn + $^{208}$Pb and $^{132}$Sn + $^{208}$Pb elastic scattering. Such a measurement
should therefore be a feasible radioactive beam experiment.

\subsection{Deformed beams on a heavy deformed target}

We have already seen in subsection \ref{proj-targ} that systems involving a deformed light nucleus incident on a deformed
heavy target lead to interesting results with effects due to strong coupling in both the projectile and the target. It
would be of interest to extend these studies to a more strongly coupled target such as $^{184}$W for instance in order
to check whether what we might term the ``greater than the sum of its parts effect'' is universal. Suitable beams for
this purpose would be $^{20}$Ne and $^{24}$Mg or $^{28}$Si. Sufficient energy
resolution to measure pure elastic scattering would require the use of a spectrometer to detect the scattered beam
particles but this ought to be possible since a detection energy resolution of about 100 keV would be necessary to 
separate the first excited state of $^{184}$W. 

\begin{figure}
\includegraphics[width=\columnwidth,clip=]{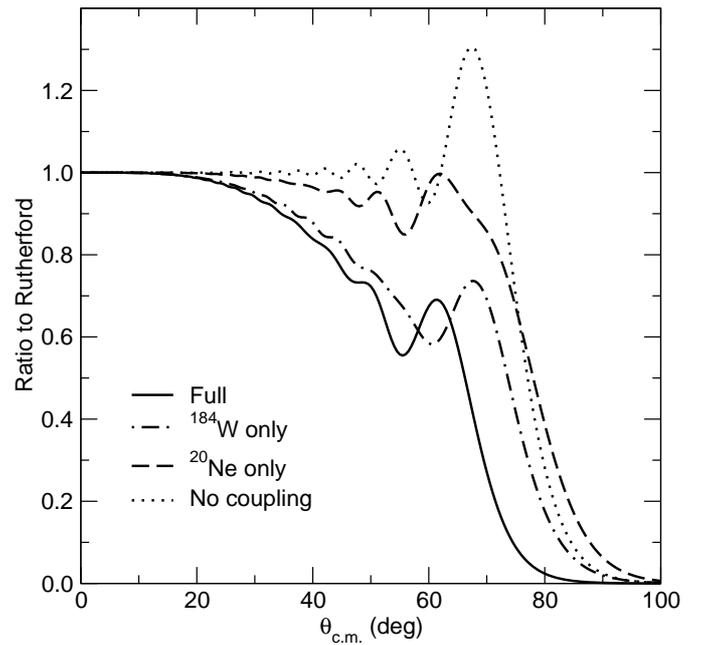}
\caption{\label{fig:ne-w}Coupled channels calculations for the $^{20}$Ne + $^{184}$W system at a bombarding energy of 120 MeV.
The dotted curve denotes the no coupling optical model calculation, the dashed curve a coupled channels calculation
including coupling to the $^{20}$Ne $2^+_1$ state only, the dot-dashed curve a coupled channels calculation including coupling
to the $^{184}$W $2^+_1$ state only and the solid curve the full coupled channels calculation including coupling to both
target and projectile $2^+_1$ states (but not mutual excitation).}
\end{figure}

In Fig.\ \ref{fig:ne-w} we plot the results of coupled channels calculations for the $^{20}$Ne + $^{184}$W system at
a bombarding energy of 120 MeV, about 23 \% higher than the nominal Coulomb barrier for this system.
The calculations again employed the Broglia-Winther prescription for the optical model parameters, the imaginary potential
depth being set at a quarter of the real depth. The $B(E2)$ values were taken from Ref.\ \cite{Ram01} and the nuclear deformation
lengths were taken from Refs.\ \cite{Gro78} and \cite{Tho77} for $^{20}$Ne and $^{184}$W respectively. In this system, unlike
the $^{20}$Ne + $^{148}$Nd data of \cite{Jia91}, the coupling effect is dominated by the target coupling (to be expected since
the coupling effect is much stronger in $^{184}$W). However, the calculations also suggest that the overall coupling effect
is not ``greater than the sum of its parts'' in this case, thus making it an interesting system for further experimental
investigation.  

\begin{figure}
\includegraphics[width=\columnwidth,clip=]{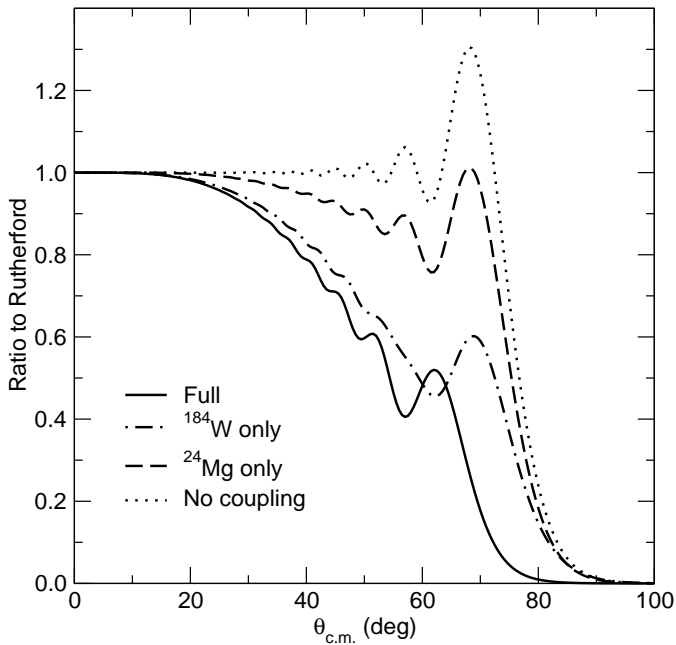}
\caption{\label{fig:mg-w}Coupled channels calculations for the $^{24}$Mg + $^{184}$W system at a bombarding energy of 145 MeV.
The dotted curve denotes the no coupling optical model calculation, the dashed curve a coupled channels calculation
including coupling to the $^{24}$Mg $2^+_1$ state only, the dot-dashed curve a coupled channels calculation including coupling
to the $^{184}$W $2^+_1$ state only and the solid curve the full coupled channels calculation including coupling to both
target and projectile $2^+_1$ states (but not mutual excitation).}
\end{figure}

As a second example we take the $^{24}$Mg + $^{184}$W system at an incident $^{24}$Mg energy of 145 MeV, again about 23 \% 
greater than the nominal Coulomb barrier for this system. Coupled channels calculations similar to those just described
for the $^{20}$Ne + $^{184}$W system are presented in Fig.\ \ref{fig:mg-w}.
The $B(E2)$ value for $^{24}$Mg was taken from Ref.\ \cite{Ram01} and the corresponding nuclear deformation length was
taken from Ref.\ \cite{Eck81}. The coupling effect is again dominated by the target coupling and seems to show similar
behaviour to $^{20}$Ne + $^{184}$W with regard to the cumulative effect of coupling to target and projectile states.

It is interesting to note that in both these cases where the target coupling effect is predicted to dominate there appears
to be a destructive interference effect between projectile and target coupling influences. This would certainly bear
further investigation and good elastic scattering data (at the $\pm$ 5 \% level) for these systems are desirable. Both experiments should
be feasible with currently available beams provided access to a spectrometer is possible in order to obtain the required detection
energy resolution of about 100 keV.

\subsection{Light weakly-bound radioactive beam on a heavy target}

While experiments to measure the near-barrier elastic scattering of light weakly-bound radioactive beams from heavy targets
have already been performed for many projectiles of interest, as outlined in section \ref{sec:RIBs}, there remain a few systems
which should prove interesting and which have not yet been studied. One such system should be $^{15}$C + $^{208}$Pb. The $^{15}$C
nucleus is relatively weakly bound, its threshold against $^{15}$C $\rightarrow$ $^{14}$C + $n$ breakup being 1.218 MeV.
However, perhaps the most important aspect of its structure is that it is a single neutron halo nucleus with the valence
neutron in a $2s_{1/2}$ orbital, thus raising the intriguing possibility of strong coupling effects due to neutron stripping
because of the long tail on the wave function of the valence neutron. Coupled reaction channels calculations have already
demonstrated that the $^{208}$Pb($^{15}$C,$^{14}$C)$^{209}$Pb single neutron stripping reaction should have a significant
coupling effect on sub-barrier $^{15}$C + $^{208}$Pb elastic scattering \cite{Kee07}. 

\begin{figure}
\includegraphics[width=\columnwidth,clip=]{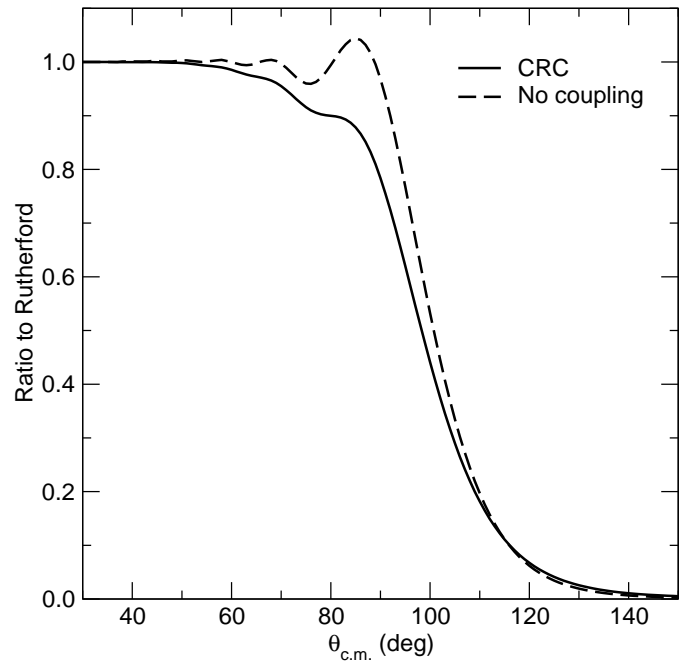}
\caption{\label{fig:c-pb}Coupled reaction channels calculation for the $^{15}$C + $^{208}$Pb system at a bombarding energy of 65 MeV.
The dashed curve denotes the no coupling optical model calculation while the solid curve denotes the coupled reaction channels
calculation including coupling to the $^{208}$Pb($^{15}$C,$^{14}$C)$^{209}$Pb single neutron stripping.}
\end{figure}

In Fig.\ \ref{fig:c-pb} we present a coupled reaction channels calculation showing the effect of coupling to the 
single neutron stripping reaction on the $^{15}$C + $^{208}$Pb elastic scattering
angular distribution for an incident $^{15}$C energy of 65 MeV. 
The entrance channel $^{15}$C + $^{208}$Pb optical potential was obtained by Watanabe-type cluster folding of $^{14}$C +
$^{208}$Pb and $n$ + $^{208}$Pb optical potentials over the internal wave function of the $^{15}$C ground state.
The $^{15}$C wave function was calculated assuming a Woods-Saxon well of ``standard'' geometry: $R_0 = 1.25 \times
{\mathrm A}^{1/3}$ fm, $a = 0.65$ fm, the well depth being adjusted to give the correct single neutron binding energy,
and a spin-orbit component of the same geometry with a fixed well depth of 6 MeV. No attempt was made to tune these
parameters. The $^{14}$C + $^{208}$Pb potential parameters were taken from a fit to the $^{12}$C + $^{208}$Pb data of Ref.\ 
\cite{San01} nearest to the appropriate energy using the potential geometry of Ref.\ \cite{Bal75} and adjusting the
real and imaginary depths. The neutron potential was the central part of pot I of Ref.\ \cite{Ann85}.  
The exit channel $^{14}$C + $^{209}$Pb potential parameters were taken from a similar fit to the appropriate data of Ref.\
\cite{San01} to those used to generate the entrance channel potential. The $\left<^{15}\mathrm{C}|^{14}\mathrm{C} + n\right>$  
and $\left<^{209}\mathrm{Pb}|^{208}\mathrm{Pb}+n\right>$ overlaps and spectroscopic factors were as in Ref.\ \cite{Kee07}.

\begin{figure}
\includegraphics[width=\columnwidth,clip=]{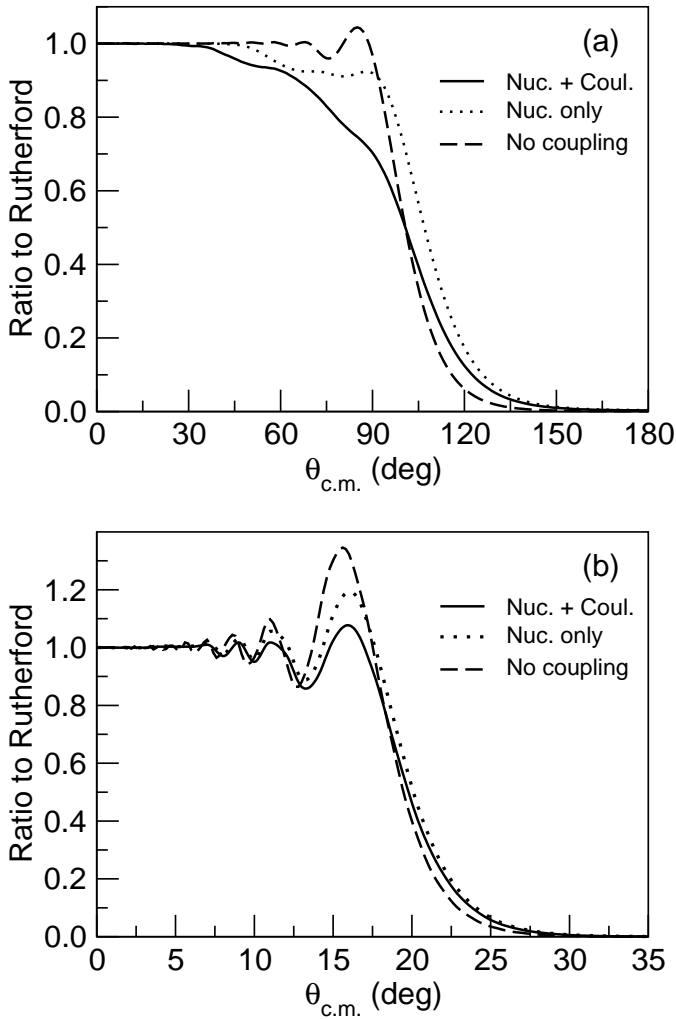}
\caption{\label{fig:15c-cdcc}(a) Coupled discretised continuum channels calculations for the $^{15}$C + $^{208}$Pb 
system at a bombarding energy of 65 MeV.
The dashed curve denotes the no coupling optical model calculation while the solid curve denotes the coupled discretised
continuum channels calculation including coupling to the $^{14}$C + $n$ continuum. The dotted curve denotes a coupled discretised
continuum channels calculation with nuclear couplings only. (b) Similar calculations for an incident $^{15}$C energy of 190 MeV,
equivalent to approximately three times the nominal Coulomb barrier for the $^{15}$C + $^{208}$Pb system. Note the different
angular ranges.}
\end{figure}

Figure \ref{fig:c-pb} shows that the single neutron stripping alone would be expected to give the characteristic ``strong
coupling'' shape to the elastic scattering angular distribution in this system. However, the weak binding of $^{15}$C
implies that breakup coupling to the continuum could produce a similar effect. In Fig.\ \ref{fig:15c-cdcc} (a) we present the
result of a coupled discretised continuum channels calculation for $^{15}$C + $^{208}$Pb at an incident energy of 65 MeV. 
The calculations are based on the same Watanabe-type folding procedure as was used to calculate the entrance channel potential
in the coupled reaction channels calculation. Couplings to the $L$ = 0, 1, 2, 3 and 4 $^{14}$C + $n$ continuum as well as the 0.74 MeV $5/2^+$
bound first excited state of $^{15}$C were included with couplings up to multipolarity $\lambda = 4$. The continuum was divided into
bins in momentum ($k$) space of width $\Delta k = 0.1$ fm$^{-1}$ up to a maximum value $k_{\mathrm{max}} = 0.6$ fm$^{-1}$. The calculated
angular distribution exhibits the strong coupling shape typical of other weakly-bound radioactive beams scattered from heavy targets. 
Also plotted on Fig.\ \ref{fig:15c-cdcc} (a), as the dotted curve, is the result of a coupled discretised continuum channels calculation
with nuclear coupling potentials only (diagonal Coulomb potentials were retained). It will be noted that the nuclear couplings are
important for $^{15}$C, suggesting that it is more similar to $^{11}$Be than to $^6$He or $^{11}$Li where the Coulomb couplings are
dominant. However, the trade off between nuclear and Coulomb couplings is to some extent energy dependent, and test calculations for
$^{11}$Li + $^{208}$Pb for an incident energy of 40 MeV yield a relative importance for nuclear and Coulomb coupling similar 
to that for 65 MeV $^{15}$C + $^{208}$Pb shown in Fig.\ \ref{fig:15c-cdcc} (a), suggesting that $^6$He may be more exceptional
than at first thought in that Coulomb breakup couplings appear to be dominant over the whole range of near-barrier energies,
see Fig.\ \ref{fig:6He} and Fig.\ 1 (c) of Ref.\ \cite{Kee13}, for example. More data (and more sophisticated calculations)
are clearly desirable to investigate further this fascinating question. 

The near barrier $^{15}$C + $^{208}$Pb elastic scattering is therefore expected to exhibit strong coupling effects 
both on account of couplings to the continuum and 
strong single neutron stripping couplings. In this respect it should be somewhat similar to the recent data for near-barrier 
$^8$He + $^{208}$Pb elastic scattering \cite{Mar13}. 
However, Fig.\ \ref{fig:15c-cdcc} (b) underlines the necessity of performing the measurements at energies close to the Coulomb
barrier. Figure \ref{fig:15c-cdcc} (b) presents the result of a similar coupled discretised continuum channels calculation to that
shown in Fig.\ \ref{fig:15c-cdcc} (a) but for an incident $^{15}$C energy of 190 MeV, approximately three times the nominal
Coulomb barrier for this system. The continuum was extended up to a maximum value $k_{\mathrm{max}} = 1.2$ fm$^{-1}$, the
bins for $k$ values above 0.6 fm$^{-1}$ being of width $\Delta k = 0.2$ fm$^{-1}$. The predicted elastic scattering angular
distribution is now seen to be of an essentially standard Fresnel type pattern with a slight residual strong Coulomb coupling
effect just before the main Coulomb-nuclear interference peak, similar to that of the 161.2 MeV $^{20}$Ne + $^{208}$Pb data
of Fig.\ \ref{fig:Ne+Pb} (b).  

\subsection{Heavier radioactive beams on a heavy target}

\begin{figure}
\includegraphics[width=\columnwidth,clip=]{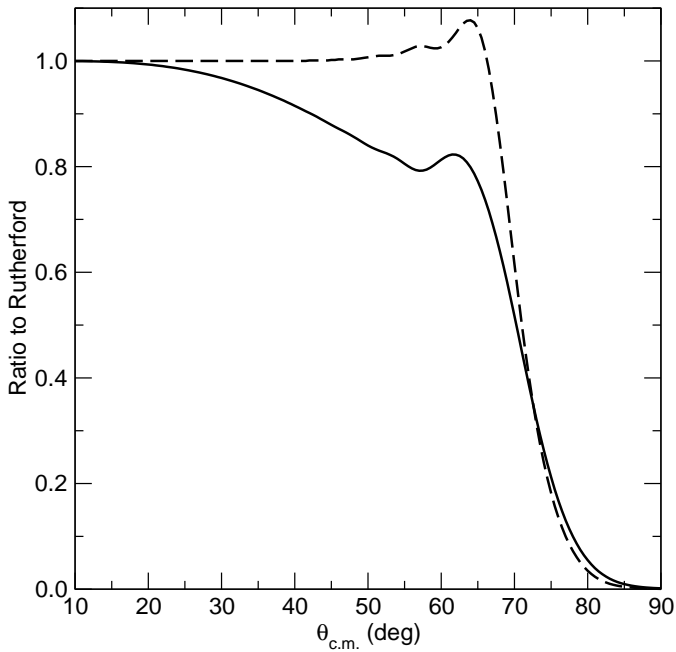}
\caption{\label{fig:30ne+pb}Coupled channels calculation for the $^{30}$Ne + $^{208}$Pb system at an incident $^{30}$Ne energy of 131 MeV.
The solid curve denotes the result of a coupled channels calculation including coupling to the $2^+_1$ state of $^{30}$Ne
only while the dashed curve denotes a no coupling optical model calculation using the same optical potential as in the 
coupled channels calculation.}
\end{figure}

Many heavier non weakly-bound radioactive nuclei are either known to be or are predicted to be more strongly deformed than the
corresponding isotopes in the valley of stability. As such they should also be interesting candidates for strong coupling
effects on their elastic scattering from heavy targets. As examples we take $^{30}$Ne and $^{32}$Mg, both with lower lying and
more strongly coupled first excited states than their already strongly coupled stable counterparts $^{20}$Ne and $^{24}$Mg.
In Fig.\ \ref{fig:30ne+pb} we present the result of a coupled channels calculation for the elastic scattering of $^{30}$Ne
from a $^{208}$Pb target at an incident $^{30}$Ne energy of 131 MeV. 
The optical model potential parameters are the same as those used in the 131 MeV $^{20}$Ne + $^{208}$Pb calculation shown in
Fig.\ \ref{fig:Ne+Pb} and the $B(E2)$ value was taken from Ref.\ \cite{Pri13}. The nuclear deformation length was derived from
the $B(E2)$ assuming the collective model expression and a charge radius of $1.2 \times {\mathrm A}^{1/3}$ fm; this value
was then multiplied by a factor of 0.75 before insertion in the coupled channels calculation since nuclear deformation
lengths derived from fits to inelastic scattering data tend to be somewhat smaller than the values derived from the
corresponding $B(E\lambda)$. As a comparison of Figs.\ \ref{fig:Ne+Pb} and \ref{fig:30ne+pb} will show, the coupling effect
for $^{30}$Ne is considerably larger than that for $^{20}$Ne at the same energy. While Fig.\ \ref{fig:30ne+pb} should not
be taken as a quantitatively reliable prediction---there may be additional effects due to strong transfer couplings and
differences in the underlying optical potential parameters for $^{30}$Ne compared to $^{20}$Ne---it does demonstrate that
$^{30}$Ne would be an interesting candidate for study when beams becomes available. 

\begin{figure}
\includegraphics[width=\columnwidth,clip=]{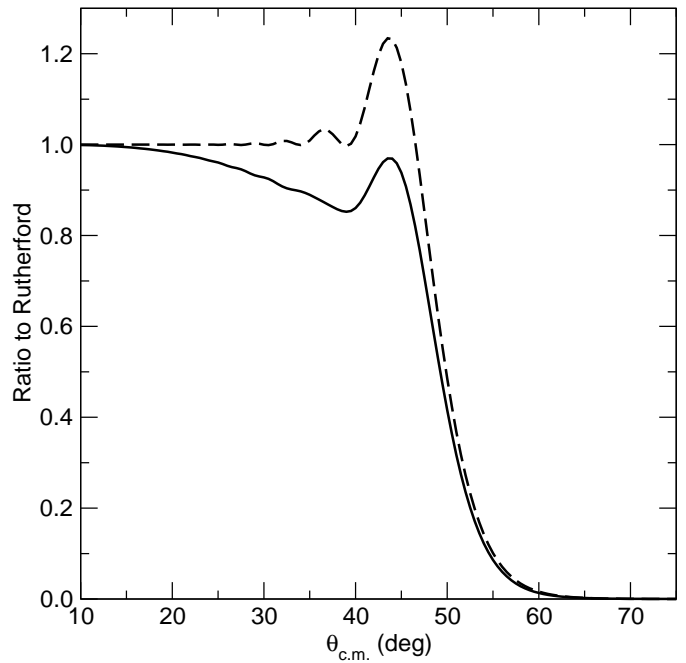}
\caption{\label{fig:32mg+pb}Coupled channels calculation for the $^{32}$Mg + $^{208}$Pb system at an incident $^{32}$Mg energy of 200 MeV.
The solid curve denotes the result of a coupled channels calculation including coupling to the $2^+_1$ state of $^{32}$Mg
only while the dashed curve denotes a no coupling optical model calculation using the same optical potential as in the
coupled channels calculation.}
\end{figure}

In Fig.\ \ref{fig:32mg+pb} we present the results of a similar calculation for the $^{32}$Mg + $^{208}$Pb system at an
incident $^{32}$Mg energy of 200 MeV. 
The optical model potential parameters were taken from a coupled channels fit to the 200 MeV $^{24}$Mg + $^{208}$Pb elastic scattering
data of Ref.\ \cite{Hen89} which included coupling to the $2^+_1$ state of $^{24}$Mg only. The $^{32}$Mg $B(E2)$ was again
taken from Ref.\ \cite{Pri13} and the nuclear deformation length derived in a similar way to the $^{30}$Ne + $^{208}$Pb calculations.
The coupling effect is considerably larger than for the $^{24}$Mg + $^{208}$Pb system at the same energy although the same caveats
apply to the quantitative reliability of the prediction as for the $^{30}$Ne + $^{208}$Pb case.

\section{Discussion}

We have shown in this review that strong coupling effects on near-barrier heavy ion elastic scattering have a long history,
going back nearly forty years. The important point to draw from the large body of work of this type conducted with stable beams
is that the elastic scattering can, under the right conditions, be sensitive to the nuclear structure of the colliding
nuclei, projectile or target or both. The advent of beams of weakly-bound light radioactive nuclei at suitable energies (and
of suitable quality) has led to a revival of interest in this topic, see for example Ref.\ \cite{Dia14} for a recent article. 

In addition to its intrinsic interest from the point of view of studies of the reaction mechanism, measurements
of the elastic scattering for systems where the
``strong coupling effect'' is apparent may have other uses. The sensitivity of the elastic scattering angular distribution 
to the details of the nuclear structure may be exploited as a means of extracting nuclear structure information, in particular 
the $B(E\lambda)$ value for the strongly coupled state. This was already suggested in the context of the application of an
analytical form of the long-ranged Coulomb DPP by Baltz et al., \cite{Bal78}: ``Indeed, in the situation where the long-range
potential arises dominantly from a single state, sub-Coulomb elastic scattering analyzed in terms of our analytical expression 
might provide an alternative method of determining the experimental $B(E2)$ to that single state.'' \cite{Bal78}.

While we would not advocate such measurements as a replacement for the traditional methods such as Coulomb excitation,
for suitable radioactive beams a measurement of the near-barrier elastic scattering from an appropriate
target might provide a valuable adjunct. The method has much to recommend it, since in the case of the $^{18}$O
+ $^{184}$W system the coupling effect on the elastic scattering was almost exclusively due to the Coulomb coupling
to the $2^+_1$ state of $^{184}$W and was thus sensitive to the $B(E2)$ value for excitation of this state; the
other members of the rotational band seemed to have little or no influence. The same was found for the $^{20}$Ne
+ $^{208}$Pb elastic scattering where the $^{20}$Ne $2^+_1$ was the relevant state; coupling to the $4^+_1$ state
had little or no effect on the elastic scattering (unlike the inelastic scattering to the $^{20}$Ne $2^+_1$). However,
practical considerations linked with the likely quality of medium to heavy mass radioactive beams will severely limit the
utility of the method, since it is essential that pure elastic scattering is measured, thus requiring good detection
energy resolution.

\begin{figure}
\includegraphics[width=\columnwidth,clip=]{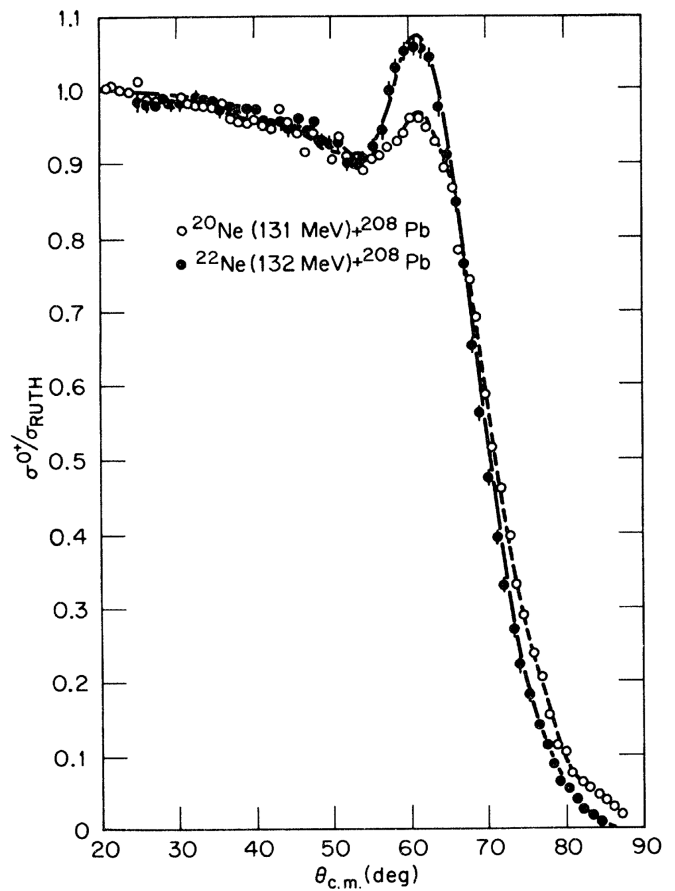}
\caption{\label{fig:2022ne}Elastic scattering angular distributions for 131 MeV $^{20}$Ne + $^{208}$Pb  and
132 MeV $^{22}$Ne + $^{208}$Pb. The solid curves denote coupled channels fits including couplings to the
$0^+_1$, $2^+_1$ and $4^+_1$ states of $^{20}$Ne and $^{22}$Ne, 
see Ref.\ \cite{Gro84} for details. Taken from Ref.\ \cite{Gro84}. Reprinted Fig.\ 1 with
permission from \cite{Gro84}. \textcircled{c} 1984, The American Physical Society}
\end{figure}

That the near-barrier elastic scattering angular distribution can, under the right circumstances, clearly show differences
due to the details of the nuclear structure of the interacting nuclei is demonstrated by Fig.\ \ref{fig:2022ne}.
The main influence on the elastic scattering is due to coupling to the $2^+_1$ state of the projectile. In $^{20}$Ne
the $2^+_1$ state has an excitation energy of 1.63 MeV and the $B(E2;0^+_1 \rightarrow 2^+_1) = 0.034$ e$^2$b$^2$ \cite{Ram01},
whereas in $^{22}$Ne the $2^+_1$ excitation energy is 1.27 MeV and the $B(E2;0^+_1 \rightarrow 2^+_1) = 0.023$ e$^2$b$^2$ \cite{Ram01}.
This difference between the two isotopes translates to an effect of order 10 \% in the measured elastic scattering angular
distributions in the region of the Coulomb-nuclear interference peak and as Fig.\ \ref{fig:2022ne} shows, the precision
required clearly to demonstrate the effect is well within the capabilities of measurements with stable beams. However, effects
of this order should also be visible in measurements with radioactive beams of the quality now becoming available and which
will be available at the second generation radioactive beam facilities when these come on line (a precision of between 2-3 \%  
in measurements of elastic scattering angular distributions has already been achieved with some currently available radioactive
beams, notably $^6$He). 

With the continuing development of existing radioactive ion beam facilities including advanced detector system updates, 
the future for these studies is bright. Further exploration of possible
strong coupling effects due to transfer reactions should be particularly propitious. However, we emphasise that care
is needed in the selection of beam and target combinations and the choice of incident energy; it is clear from the data and
calculations presented in this review that a beam energy close to the nominal Coulomb barrier for the system under study
is essential if strong coupling effects on the elastic scattering are to be observed unambiguously. The one exception to this
rule seems to be $^{11}$Li, where the coupling to the continuum is so strong that it should persist to energies well in
excess of the Coulomb barrier, at least for heavy targets. On the other hand, in $^8$B we have the intriguing case of a nucleus
which ought to be exotic---it has the lowest known threshold against breakup and the predicted breakup cross sections are very
large---and yet is not, at least not in the conventional sense, since coupling to the $^7$Be + $p$ continuum has an almost
negligible effect in coupled discretised continuum channels calculations. Why this should be so remains ``a riddle, 
wrapped in a mystery, inside an enigma'' \cite{Chu39}. One might almost say that $^8$B is exotic because it is {\em not} exotic \ldots
Nevertheless, it is clear that much work remains to be done in this fascinating field, proving that elastic scattering measurements
are neither trivial nor uninteresting. 
\section*{Acknowledgements}
The authors would like to thank Dr.\ D.D. Caussyn for help with scanning figures for this article.
KWK acknowledges partial support from the Florida State University Robert O. Lawton Fund.
Figures 8 and 10 are \textcircled{c} 1984 and 1992 respectively, reprinted with permission from Elsevier.

\end{document}